\documentclass[%
 reprint,
 superscriptaddress,
 amsmath,amssymb,
 aps,
prb,
]{revtex4-2}

\usepackage{graphicx}% Include figure files
\usepackage{dcolumn}% Align table columns on decimal point
\usepackage{bm}% bold math
\usepackage{etoolbox}
\usepackage{setspace}
\usepackage{ulem}
\usepackage{physics}

\usepackage[mathlines]{lineno}% Enable numbering of text and display math
\usepackage{xspace}
\usepackage{comment}

\newcommand{\figref}[2]{\hyperref[#1]{Fig.~\ref{#1}{#2}}}
\newcommand{\figureref}[2]{\hyperref[#1]{Figure~\ref{#1}{#2}}}
\newcommand{\solofigureref}[2]{\hyperref[#1]{Figure~\ref{#1}}}

\renewcommand{\sout}[1]{}
\usepackage{hyperref}
\hypersetup{colorlinks,allcolors=blue}

\begin{document}

\title{Direct Observation of the Lindhard Continuum using Resonant Inelastic X-ray Scattering}

\author{Eder~G.~Lomeli}
\affiliation{Department of Materials Science and Engineering, Stanford University, Stanford, CA, USA}
\affiliation{Stanford Institute for Materials and Energy Sciences, SLAC National Accelerator Laboratory, Menlo Park, CA, USA}

\author{Sarbajaya~Kundu}
\affiliation{Department of Physics and Astronomy, University of Notre Dame, Notre Dame, IN, USA}
\affiliation{Stavropoulos Center for Complex Quantum Matter, University of Notre Dame, Notre Dame, IN, USA}

\author{Yi-De~Chuang}
\affiliation{Advanced Light Source, Lawrence Berkeley National Laboratory, 1 Cyclotron Road, Berkeley, CA, USA}

\author{Zengqing~Zhuo}
\affiliation{Advanced Light Source, Lawrence Berkeley National Laboratory, 1 Cyclotron Road, Berkeley, CA, USA}

\author{Ke~Chen}
\affiliation{Department of Physics, Temple University, Philadelphia, PA, USA}

\author{Xiaoxing~Xi}
\affiliation{Department of Physics, Temple University, Philadelphia, PA, USA}

\author{Lingjia~Shen}
\affiliation{Linac Coherent Light Source, SLAC National Accelerator Laboratory, Menlo Park, CA, USA}

\author{Georgi~Dakovski}
\affiliation{Linac Coherent Light Source, SLAC National Accelerator Laboratory, Menlo Park, CA, USA}

\author{Stephan~Gepr\"ags}
\affiliation{Walther-Meissner-Institut, Bayerische Akademie der Wissenschaften, Garching, Germany}

\author{Brian~Moritz}
\affiliation{Stanford Institute for Materials and Energy Sciences, SLAC National Accelerator Laboratory, Menlo Park, CA, USA}

\author{Thomas~P.~Devereaux}
\affiliation{Department of Materials Science and Engineering, Stanford University, Stanford, CA, USA}
\affiliation{Stanford Institute for Materials and Energy Sciences, SLAC National Accelerator Laboratory, Menlo Park, CA, USA}

\author{John~Vinson}
\affiliation{Material Measurement Laboratory, National Institute of Standards and Technology, Gaithersburg, MD, USA}

\author{Matthias~F.~Kling}
\affiliation{Linac Coherent Light Source, SLAC National Accelerator Laboratory, Menlo Park, CA, USA}
\affiliation{Department of Applied Physics, Stanford University, Stanford, CA, USA}

\author{Edwin~W.~Huang}
\email{ehuang3@nd.edu}
\affiliation{Department of Physics and Astronomy, University of Notre Dame, Notre Dame, IN, USA}
\affiliation{Stavropoulos Center for Complex Quantum Matter, University of Notre Dame, Notre Dame, IN, USA}

\author{Daniel~Jost}
\email{daniel.jost@stanford.edu}
\affiliation{Linac Coherent Light Source, SLAC National Accelerator Laboratory, Menlo Park, CA, USA}

\date{\today}% It is always \today, today,
             %  but any date may be explicitly specified

\begin{abstract}
Understanding the excitations of quantum materials is essential for unraveling how their microscopic constituents interact. Among these, particle-hole excitations form a particularly important class, as they govern fundamental processes such as screening, dissipation, and transport. In metals, the continuum of electron-hole excitations is described by the Lindhard function~\cite{Lindhard1954}. Although central to the theory of Fermi liquids, the corresponding Lindhard continuum has remained experimentally elusive. Here, we report its direct observation in the weakly correlated metal MgB$_{2}$ using ultra-soft resonant inelastic X-ray scattering (RIXS). We resolve a linearly dispersing excitation with velocity comparable to the Fermi velocity and find quantitative agreement with simulations of the non-interacting charge susceptibility. A detailed analysis and decomposition of the simulations reveal the intra-band origin of this low-energy excitation, confirming it as the Lindhard continuum. Our results establish ultra-soft RIXS as a momentum-resolved probe of the fermiology in metals and call for deeper investigations of continuum features in RIXS and related spectroscopy of other materials beyond MgB$_{2}$.
\end{abstract}

\maketitle

\begin{figure*}
    \centering
    \includegraphics[width=140mm]{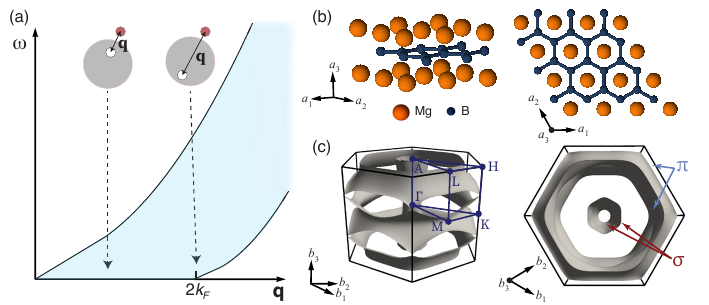}
    \caption{(a) The range of momentum $\mathbf{q}$ and the corresponding energy $\omega$ for which particle-hole excitations, taking a particle below the Fermi surface to a higher energy above it, can be created in a Fermi liquid. (b) MgB$_{2}$, a weakly interacting Fermi liquid system, exhibits a hexagonal structure, with 2D layers of B atoms sandwiching triangular Mg layers. (c)  The Fermi surfaces of MgB$_{2}$ consist of a 3D tubular network of mostly boron $\pi$ states and 2D cylindrical sheets derived mostly from boron $\sigma$ states.}
    \label{fig:Fig1_intro}
\end{figure*}

The most elementary low-energy excitations of a Fermi liquid are particle-hole excitations obtained by scattering a quasiparticle across the Fermi surface. In the absence of strong interactions, Fermi liquid theory predicts these excitations form a continuous, bounded region in energy and momentum space known as the Lindhard continuum (\figref{fig:Fig1_intro}a)~\cite{Lindhard1954}. This continuum of excitations is the starting point from which to understand collective excitations, including plasmons, acoustic plasmons, and sound modes. Despite the fundamental role and importance of the Lindhard continuum, its experimental signatures have only been indirect, for instance, in contributing to the linewidth of plasmons via Landau damping. While related inter-band particle-hole excitations are visible even by optical spectroscopy, the low-energy intra-band excitations have mysteriously eluded spectroscopic observation. The Lindhard continuum is encoded in the dynamical charge susceptibility, and probes such as electron energy loss spectroscopy (EELS) have, in principle, the correct energy and momentum range to detect it. Yet surprisingly, no feature resembling the textbook illustration of \figref{fig:Fig1_intro}a has been reported to date.

As we shall discuss, the limitation of EELS in probing the Lindhard continuum arises from that fact that EELS measures the screened longitudinal charge response $-\Im \epsilon^{-1}(q,\omega)$. In metals at small $q$, charge conservation together with the $\frac{1}{q^2}$ Coulomb interaction and static screening concentrates spectral weight into the plasmon, leaving the intra-band particle-hole continuum with vanishing intensity. By contrast, resonant inelastic X-ray scattering (RIXS) couples to local, orbital-selective operators that are not tied to a conserved density and not dressed by the long-range Coulomb kernel. As a result, RIXS is free from the screening constraint, and with the appropriate material, resonance condition, and experimental geometry, RIXS is well poised to observe the Lindhard continuum.

\begin{figure*}
    \centering
    \includegraphics[width=180mm]{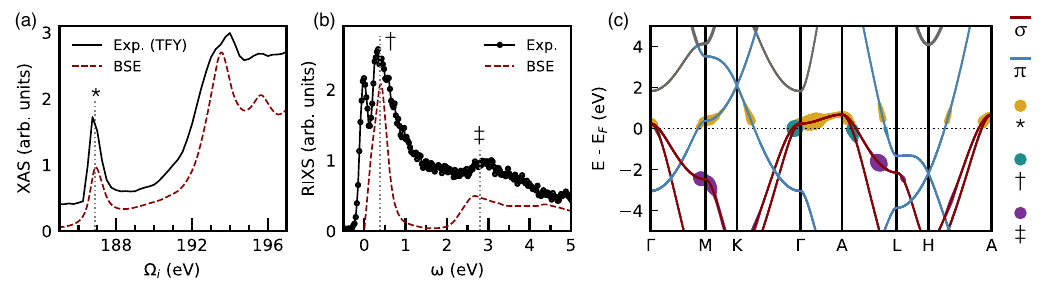}
    \caption{(a) Total Florescence Yield (TFY) XAS measurement with corresponding Bethe-Salpeter equation (BSE) calculation. The pre-edge feature of interest is highlighted (*). (b) RIXS ($q_\parallel =0.016$~\AA$^{-1}$) measurement of the pre-edge feature with corresponding BSE calculation. (c) Band structure of MgB$_2$ denoting the $\sigma$ and $\pi$ bands, as well as the projected XAS ($*$) and RIXS ($\dagger$ and $\ddagger$) transitions onto the relevant DFT eigenstates. $E_F$ is marked by the horizontal gray dotted line.}
    \label{fig:Fig2}
\end{figure*}

The material we investigate here using RIXS is the weakly correlated metal MgB$_{2}$. It is best known for its superconductivity below 39\,K~\cite{Nagamatsu:2001,Kang:2001}, the highest transition temperature among ambient pressure conventional superconductors, and for its unusual superconducting properties~\cite{Chen:2001,Choi2002,Souma2003}. More recently, its potential for hosting topological properties has attracted attention~\cite{Jin:2019,Zhou2019,An:2001,Yu2024}. MgB$_{2}$ consists of two-dimensional honeycomb layers of boron atoms sandwiched by triangular Mg layers (\figref{fig:Fig1_intro}{b}). Its electronic structure is well described by density functional theory (DFT) and exhibits minimal electronic correlations~\cite{Souma2003,Mazin:2003,Silkin:2009,Sharma2011}. The Fermi surfaces consist of two $\sigma$ bands of boron $2s,2p_{x,y}$ orbital character forming cylinders along the $k_z$ direction at the Brillouin zone (BZ) center and two $\pi$ bands of boron $2p_{z}$ orbital character spread along the BZ boundary (\figref{fig:Fig1_intro}{c}). Since the boron K edge involves $1s\to 2p$ transitions, RIXS at this edge probes excitations arising from these bands with both orbital and momentum specificity. Crucially, the experimental geometry enables precise control over the in-plane momentum transfer in the regime below $0.2$~\AA$^{-1}$. Given that the $\sigma$ bands have Fermi velocities of roughly 2 eV\AA, their intra-band excitations would have energies of a couple hundred meV and if visible, would be resolvable with our experimental resolution of $\approx 100$~meV.

To probe the possible resonances available to us for RIXS, we first perform X-ray absorption spectroscopy (XAS) at the B K edge. The spectrum exhibits a sharp feature near 187 eV [\figref{fig:Fig2}{a}], corresponding to transitions from the B $1s$ core level into low lying unoccupied $2p$ states, as determined from Bethe-Salpeter equation (BSE) calculations based on Density Functional Theory (DFT) wavefunctions (see Methods)~\cite{Vinson_BSE,Vinson_OCEAN3,Giannozzi_QEspresso,Hamann_ONCVPSP,Sun_SCAN}. Our experimental spectrum resembles previous XAS measurements, where, based on local density of states calculations, the pre-edge feature was mainly associated with planar $p_x,p_y$ hole states near the Fermi surface~\cite{Zhu:2002, Saito:2012}. Looking at the band structure [\figref{fig:Fig2}{c} or \figref{fig:all_bs_suppl}], this feature corresponds to both unoccupied $\sigma$ band states near $\Gamma$/$A$ and, with a lesser contribution, unoccupied $\pi$ band states near $M$. This resonance selectively enhances excitations involving $\sigma$ and $\pi$ states near $E_F$ and thus serves as the optimal incident energy for probing low energy $intra$-band dynamics using RIXS.

 \begin{figure*}
    \centering
    \includegraphics[width=180mm]{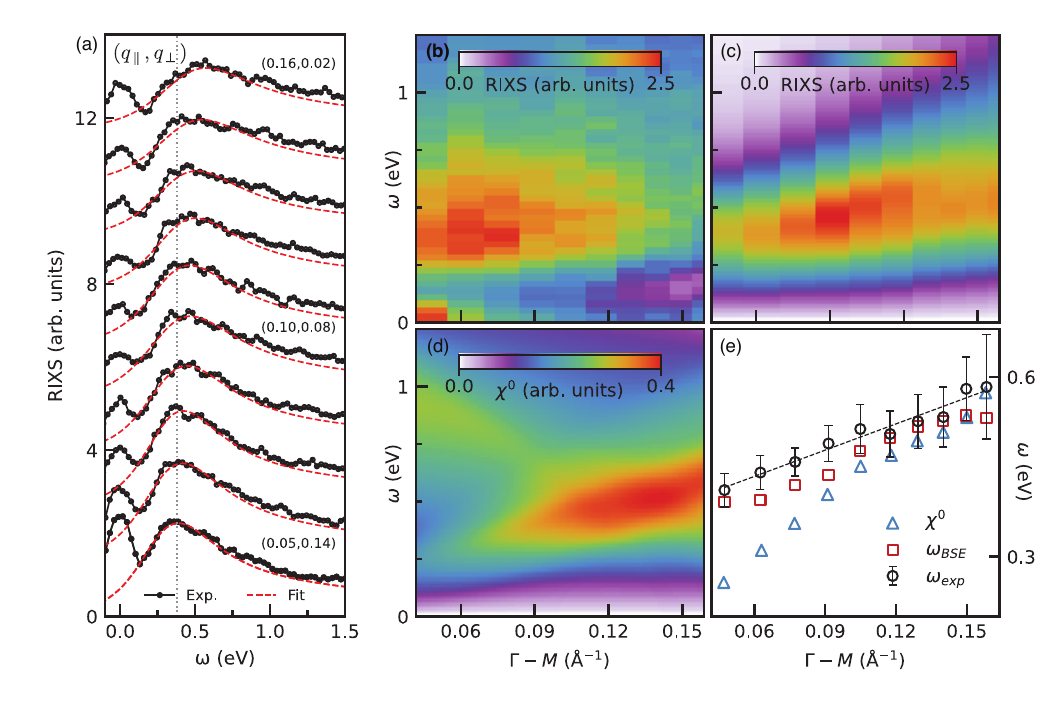}
    \caption{(a) RIXS intensity linecuts showing the out-of-plane and in-plane dependence of the low-energy feature. Dotted gray line highlights the peak position of the low energy voigt-fit component at the lowest $q_\parallel$, for comparison at higher values. (b), (c) Experimental and simulated RIXS color maps depicting the momentum dependence of the low-energy excitation. (d) Lindhard function $\chi^0(q,\omega)$ over the relevant range of energy $\omega$ and in-plane momentum $q_{\Vert}$, broadened by experimental energy resolution of 100 meV. (e) Dispersion of the low energy features across experiment and calculations of RIXS and $\chi^0$ simulations.}
    \label{fig:Fig3}
\end{figure*}

The RIXS spectra recorded at an incident energy of 187 eV exhibit two distinct excitation regimes (\figref{fig:Fig2}{b}). At energy transfers above 2 eV, we observe broad and momentum-independent features consistent with inter-band transitions between the $\sigma$ and $\pi$ manifolds, shaded red and blue in \figref{fig:Fig2}{c}, respectively. Similar features have been reported previously by EELS measurements \cite{Idrobo:2004}. Closer to the elastic intensity, a strong feature is visible
%that corresponds to solely \textit{intra}-band transitions within both $\sigma$ or $\pi$ bands, with a breakdown of each contribution discussed later [see \figureref{fig:fig4}]. 
%This excitation appears 
in a narrow energy window of a few hundred meV. This feature exhibits a markedly asymmetric shape (\figref{fig:Fig3}{a}), and to the best of our knowledge, has not been reported in prior experiments. Projecting the intermediate state responsible for the low-energy feature in our BSE calculation onto the band structure reveals that the transition is localized within the $\sigma$ bands of the Fermi surface close to the BZ center (see \figref{fig:Fig2}{c}), suggesting a possible intraband nature of the transition. Indeed, it lies in the correct energy range for intra-band excitations, according to \textit{ab initio} calculations of the dielectric function\cite{Zhukov:2001}. Interestingly, these same calculations predict that the loss function, as would be measured in an EELS experiment, is dominated by a plasma mode near 3 eV and has essentially no intensity at lower energies. 

To further characterize this low energy feature, we vary the incident angle $\theta_\mathrm{in}$, thereby tuning the in-plane momentum $q_\parallel$ transfer up to $\approx 0.17$ \AA$^{-1}$. As shown in \figref{fig:Fig3}{a} and \figref{fig:Fig3}{b}, the low-energy feature shifts systematically with increasing $q_\parallel$, indicating a dispersive character. This dispersion is further supported by the extracted peak positions (\figref{fig:Fig3}{e}) from Voigt fits to our data (see example in \figref{fig:voigt_fit}). The low-energy excitation follows a nearly linear dispersion across the accessible momentum range, a behavior consistent with intra-band particle-hole excitations in a low-energy dispersive continuum. Its speed of $\approx 1.4$ eV \AA, around half of the Fermi velocity of the $\sigma$ bands, strongly suggests a connection to the low-energy band structure.

\begin{figure*}
\centering
\includegraphics[width=180mm]{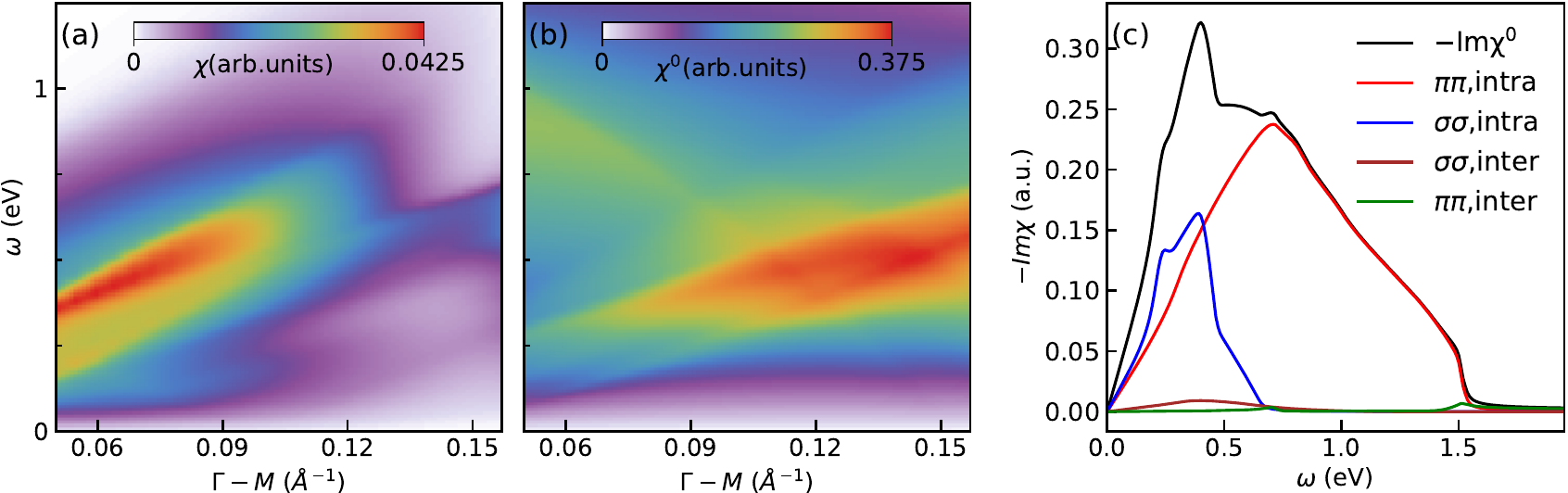}
\caption{(a,b) RPA charge susceptibility $\chi$, weighed by sinusoidal factors accounting for the experimental geometry, and Lindhard function $\chi^0$ over the relevant range of energy $\omega$ and in-plane momentum $q_{\Vert}$. In contrast to \figref{fig:Fig3}{d}, no broadening is applied. (c) Breakdown of the non-interacting charge susceptibility into its intraband and interband components at $q_{\Vert}=0.088$ \AA$^{-1}$. Inter-band excitations between the $\sigma$ and $\pi$ bands are negligible.}
\label{fig:fig4}
\end{figure*}

Further BSE-based simulations of the RIXS spectra (\figref{fig:Fig3}{c}), incorporating the full experimental geometry, help us to interpret these trends. The calculations reproduce both the dispersion and overall intensity variation observed experimentally, including the suppression of spectral weight at grazing conditions of the X-rays. In the experiment, this suppression stems from increasing out-of-plane components of the X-ray polarization that naturally decreases the interaction with in-plane $\sigma-$type states. Similarly, in our calculation, the polarization is accounted for when calculating the cross section of the valence-hole--conduction-electron excitation, reproducing an intensity profile consistent with experiment.

We additionally compare the experimental data against a calculation of the Lindhard function based on an eight-orbital tight-binding model given in \ref{tbm} \cite{Yu2024}. The Lindhard function, which consists purely of elementary electron-hole excitations, is shown in \figref{fig:Fig3}{d}. As the Lindhard function does not account for the experimental geometry and orbital sensitivity, its intensity is peaked at a larger value of $q$. Yet remarkably, its dispersion [\figref{fig:Fig3}{e}] matches that of the experiment closely except at smaller $q$ where the spectral weight of the Lindhard function is split between different features. The overall agreement between the simulations and experimental data strongly suggest that elementary electron-hole excitations are responsible for the low energy feature seen in RIXS.

The Lindhard calculations can be extended by considering interaction effects and the angle-dependent orbital sensitivity of the experiment. Under the random phase approximation (RPA), we compute the the charge susceptibility as a matrix, separating the response due to different orbitals. The contributions from the $p_y$ and $p_z$ orbitals are weighted according to the incident angle $\theta_\mathrm{in}$ and summed to approximate the RIXS process's geometry-dependent orbital selectivity. The result of this orbitally-resolved RPA calculation is shown in \figref{fig:fig4}{a}. In comparing against the non-interacting susceptibility (i.e.~Lindhard function), now plotted in  \figref{fig:fig4}{b} without broadening, we see that the combination of interactions and orbital sensitivity shifts the spectral weight between the various dispersive features. Indeed, accounting for the experimental geometry in this minimal fashion, the intensity peak has an energy and momentum that matches the RIXS data.

The close agreement between calculations and experiment motivates a deeper analysis of the simulations to pinpoint the microscopic origin of the low-energy excitation. The low-energy spectral weight of the BSE spectra shown in \figref{fig:Fig2}{b} and \figref{fig:Fig3}{c} can be analyzed by projecting onto the band structure. The result is indicated by the thickness of the curves in \figref{fig:Fig2}{c}, showing that $\sigma$ states near $E_F$ are the most relevant contributors to the low energy spectral weight. However, considering that there are two $\sigma$ bands, additional analysis is needed to delineate whether the feature is due to intra- or inter-band excitations.

For this purpose, we exploit the fact that the Lindhard function can be separated into components containing inter-band and intra-band excitations. The result of this dissection is shown in \figref{fig:fig4}{c}. The overwhelming majority of the spectral weight below 1eV is due to intra-band excitations, with a significant fraction from both the two $\sigma$ bands and two $\pi$ bands. Inter-band excitations play a much more minor role in this energy range. Given the overall similarity of these calculations with both the BSE-based RIXS simulation and the experimental spectra, this analysis conclusively demonstrates that the features around 0.5 eV are due to intra-band excitations. In other words, our experimental spectra in \figref{fig:Fig3}{b} directly map out the Lindhard continuum.

Our observation of the Lindhard continuum by RIXS raises the question of whether other experimental probes are similarly capable of directly observing the continuum. The standard textbook discussion of the susceptibility of the Fermi liquid teaches that with long-range Coulomb interactions, the susceptibility at small $q$ contains both the continuum and a plasma peak. What is seldom mentioned is that the the continuum has exceedingly small intensity. This can be understood by considering two fundamental sum rules for metallic systems
\begin{equation}
    \int_0^\infty \omega L(q, \omega) d\omega = \frac{\pi}{2} \omega_p^2\quad,\quad
    \int_0^\infty \frac{L(q, \omega)}{\omega} d\omega = \frac{\pi}{2},
\end{equation}
where $L(q,\omega)$ = $-\Im \epsilon^{-1}(q,\omega) = - v(q) \Im \chi(q,\omega)$ is the loss function. The first equation is the $f-$sum rule. The second is due to static screening $\epsilon(q, 0) \to \infty$ and applying the Kramers-Kronig relations to $v(q) \chi(q,\omega) = \epsilon^{-1}(q,\omega) - 1$.
At small $q$, the $f-$sum rule requires $L(q,\omega) = \frac{\pi}{2} \omega_P \delta(\omega - \omega_P) + L_{p-h}(q,\omega)$, where $\omega_P$ is the plasma frequency. The latter term, representing the continuum, has no contribution to the first moment because it is kinematically constrained to $\omega \leq v_F q$. Inserting this into the screening sum rule, it is evident that the plasmon alone saturates the sum rule, leaving no intensity for $L_{p-h}(q,\omega)$.

This sum rule analysis demonstrates that EELS or non-resonant inelastic X-ray scattering, which probe the total charge response, cannot see low-energy intra-band excitations. RIXS, which instead probes an orbitally-selective projected charge density, does not suffer from the constraints imposed by the screening sum rule and will generically capture the Lindhard continuum. In \ref{orbitalselect}, a simplified calculation under the RPA shows that orbitally-selected susceptibilities contain intra-band excitations with order 1 intensity whereas the total charge susceptibility has a $q^4$ suppression.

Our analysis holds beyond the RPA, since the relevant sum rules constrain even strongly interacting systems. Whenever a spectral feature is confined to $\omega \leq v_F q$, the screening rule for the longitudinal charge response severely limits its intensity at small $q$. By contrast, higher energy continua, such as the strange metal continuum reported in \cite{Mitrano:2018}, are not bound by the sum rules; so their visibility in EELS is consistent with our framework.

By combining elemental and orbital selectivity with ultra-small momentum transfer, RIXS provides direct access to the elementary excitations of a Fermi liquid and the fermiology governed by Fermi surface geometry and low energy band structure. This capability bridges a long-standing gap between photoemission (single-particle spectra) and transport (long-wavelength charge dynamics) and delivers, to our knowledge, unprecedented momentum space granularity in tracking intra-band particle-hole excitations in a metal.

While our demonstration is in a canonical Fermi liquid, the same combination of orbital selectivity and low-$q$ sensitivity suggests that RIXS can become a powerful probe of elementary charge dynamics in more exotic regimes of strongly correlated systems such as strange metals. Our results therefore motive systematic low-$q$, polarization-resolved RIXS studies across families of weakly and strongly correlated materials, and a re-examination of features often treated as ``background''. In systems where no general analog of the Lindhard continuum is established, the presence, lineshape, and $q, \omega$ scaling of such continua would provide sharp, falsifiable tests of candidate theories of correlated metals.

\begin{acknowledgments}
XAS and RIXS experiments used resources from the Advanced Light Source, a U.S. DOE Office of Science User Facility under contract no. DE-AC02-05CH11231. The scientists at the Linac Coherent Light Source (LCLS), SLAC National Accelerator Laboratory, were supported by the U.S. Department of Energy~(DOE), Office of Science, Office of Basic Energy Sciences~(BES) under contract no.~DE-AC02-76SF00515. 

Theoretical work for the {\sc Quantum ESPRESSO} and {\sc ocean} calculations (E.G.L, B.M., and T.P.D) was supported in part by the U.S. Department of Energy, Office of Basic Energy Sciences, Materials Sciences and Engineering Division. Calculations were performed on resources of the Center for Research Computing at the University of Notre Dame, on the Sherlock cluster at Stanford University, and on resources of the National Energy Research Scientific Computing Center (NERSC), a Department of Energy Office of Science User Facility, using NERSC award BES-ERCAP0031424. 

MgB$_2$ sample preparation at Temple University was supported by the grant DE-SC0022330 funded by the U.S. Department of Energy, Office of Science. 

\subsection{Declarations}
Certain equipment, instruments, software, or materials are identified in this paper in order to specify the experimental procedure adequately.  Such identification is not intended to imply recommendation or endorsement of any product or service by NIST, nor is it intended to imply that the materials or equipment identified are necessarily the best available for the purpose.

\subsection{Author Contributions}
D.J. conceived the study. K.C. and X.X. synthesized the samples. E.G.L. and D.J. performed the XAS and RIXS measurements with support from Y.D. and Z.Z. E.G.L. and J.V. conducted the {\sc Quantum ESPRESSO} and {\sc ocean} calculations of XAS and RIXS with support from B.M. and T.P.D. S.K. and E.W.H. developed the theory for interpreting the results and conducted the Lindhard function and RPA calculations. E.G.L., S.K., E.W.H., and D.J. prepared the initial version of the manuscript. All authors contributed to the discussion of the results and the final version of the manuscript.

\subsection{Data Availability}
Data is available from authors upon reasonable request.

\subsection{Competing Interests}
The authors declare no competing interests.

\end{acknowledgments}

\newpage

\bibliography{MgB2_bib}

%apsrev4-2.bst 2019-01-14 (MD) hand-edited version of apsrev4-1.bst
%Control: key (0)
%Control: author (8) initials jnrlst
%Control: editor formatted (1) identically to author
%Control: production of article title (0) allowed
%Control: page (0) single
%Control: year (1) truncated
%Control: production of eprint (0) enabled
\begin{thebibliography}{31}%
\makeatletter
\providecommand \@ifxundefined [1]{%
 \@ifx{#1\undefined}
}%
\providecommand \@ifnum [1]{%
 \ifnum #1\expandafter \@firstoftwo
 \else \expandafter \@secondoftwo
 \fi
}%
\providecommand \@ifx [1]{%
 \ifx #1\expandafter \@firstoftwo
 \else \expandafter \@secondoftwo
 \fi
}%
\providecommand \natexlab [1]{#1}%
\providecommand \enquote  [1]{``#1''}%
\providecommand \bibnamefont  [1]{#1}%
\providecommand \bibfnamefont [1]{#1}%
\providecommand \citenamefont [1]{#1}%
\providecommand \href@noop [0]{\@secondoftwo}%
\providecommand \href [0]{\begingroup \@sanitize@url \@href}%
\providecommand \@href[1]{\@@startlink{#1}\@@href}%
\providecommand \@@href[1]{\endgroup#1\@@endlink}%
\providecommand \@sanitize@url [0]{\catcode `\\12\catcode `\$12\catcode
  `\&12\catcode `\#12\catcode `\^12\catcode `\_12\catcode `\%12\relax}%
\providecommand \@@startlink[1]{}%
\providecommand \@@endlink[0]{}%
\providecommand \url  [0]{\begingroup\@sanitize@url \@url }%
\providecommand \@url [1]{\endgroup\@href {#1}{\urlprefix }}%
\providecommand \urlprefix  [0]{URL }%
\providecommand \Eprint [0]{\href }%
\providecommand \doibase [0]{https://doi.org/}%
\providecommand \selectlanguage [0]{\@gobble}%
\providecommand \bibinfo  [0]{\@secondoftwo}%
\providecommand \bibfield  [0]{\@secondoftwo}%
\providecommand \translation [1]{[#1]}%
\providecommand \BibitemOpen [0]{}%
\providecommand \bibitemStop [0]{}%
\providecommand \bibitemNoStop [0]{.\EOS\space}%
\providecommand \EOS [0]{\spacefactor3000\relax}%
\providecommand \BibitemShut  [1]{\csname bibitem#1\endcsname}%
\let\auto@bib@innerbib\@empty
%</preamble>
\bibitem [{\citenamefont {Lindhard}(1954)}]{Lindhard1954}%
  \BibitemOpen
  \bibfield  {author} {\bibinfo {author} {\bibfnamefont {J.}~\bibnamefont
  {Lindhard}},\ }\bibfield  {title} {\bibinfo {title} {On the properties of a
  gas of charged particles},\ }\href@noop {} {\bibfield  {journal} {\bibinfo
  {journal} {Danske Matematisk-fysiske Meddelelser}\ }\textbf {\bibinfo
  {volume} {28}},\ \bibinfo {pages} {1} (\bibinfo {year} {1954})}\BibitemShut
  {NoStop}%
\bibitem [{\citenamefont {Nagamatsu}\ \emph {et~al.}(2001)\citenamefont
  {Nagamatsu}, \citenamefont {Nakagawa}, \citenamefont {Muranaka},
  \citenamefont {Zenitani},\ and\ \citenamefont {Akimitsu}}]{Nagamatsu:2001}%
  \BibitemOpen
  \bibfield  {author} {\bibinfo {author} {\bibfnamefont {J.}~\bibnamefont
  {Nagamatsu}}, \bibinfo {author} {\bibfnamefont {N.}~\bibnamefont {Nakagawa}},
  \bibinfo {author} {\bibfnamefont {T.}~\bibnamefont {Muranaka}}, \bibinfo
  {author} {\bibfnamefont {Y.}~\bibnamefont {Zenitani}},\ and\ \bibinfo
  {author} {\bibfnamefont {J.}~\bibnamefont {Akimitsu}},\ }\bibfield  {title}
  {\bibinfo {title} {{Superconductivity at 39{\thinspace}K in magnesium
  diboride}},\ }\href {https://doi.org/10.1038/35065039} {\bibfield  {journal}
  {\bibinfo  {journal} {Nature}\ }\textbf {\bibinfo {volume} {410}},\ \bibinfo
  {pages} {63} (\bibinfo {year} {2001})}\BibitemShut {NoStop}%
\bibitem [{\citenamefont {Kang}\ \emph {et~al.}(2001)\citenamefont {Kang},
  \citenamefont {Kim}, \citenamefont {Choi}, \citenamefont {Jung},\ and\
  \citenamefont {Lee}}]{Kang:2001}%
  \BibitemOpen
  \bibfield  {author} {\bibinfo {author} {\bibfnamefont {W.~N.}\ \bibnamefont
  {Kang}}, \bibinfo {author} {\bibfnamefont {H.-J.}\ \bibnamefont {Kim}},
  \bibinfo {author} {\bibfnamefont {E.-M.}\ \bibnamefont {Choi}}, \bibinfo
  {author} {\bibfnamefont {C.~U.}\ \bibnamefont {Jung}},\ and\ \bibinfo
  {author} {\bibfnamefont {S.-I.}\ \bibnamefont {Lee}},\ }\bibfield  {title}
  {\bibinfo {title} {{MgB$_2$} superconducting thin films with a transition
  temperature of 39 {Kelvin}},\ }\href
  {https://doi.org/10.1126/science.1060822} {\bibfield  {journal} {\bibinfo
  {journal} {Science}\ }\textbf {\bibinfo {volume} {292}},\ \bibinfo {pages}
  {1521} (\bibinfo {year} {2001})}\BibitemShut {NoStop}%
\bibitem [{\citenamefont {Chen}\ \emph {et~al.}(2001)\citenamefont {Chen},
  \citenamefont {Konstantinovi\ifmmode~\acute{c}\else \'{c}\fi{}},
  \citenamefont {Irwin}, \citenamefont {Lawrie},\ and\ \citenamefont
  {Franck}}]{Chen:2001}%
  \BibitemOpen
  \bibfield  {author} {\bibinfo {author} {\bibfnamefont {X.~K.}\ \bibnamefont
  {Chen}}, \bibinfo {author} {\bibfnamefont {M.~J.}\ \bibnamefont
  {Konstantinovi\ifmmode~\acute{c}\else \'{c}\fi{}}}, \bibinfo {author}
  {\bibfnamefont {J.~C.}\ \bibnamefont {Irwin}}, \bibinfo {author}
  {\bibfnamefont {D.~D.}\ \bibnamefont {Lawrie}},\ and\ \bibinfo {author}
  {\bibfnamefont {J.~P.}\ \bibnamefont {Franck}},\ }\bibfield  {title}
  {\bibinfo {title} {Evidence for two superconducting gaps in {MgB$_2$}},\
  }\href {https://doi.org/10.1103/PhysRevLett.87.157002} {\bibfield  {journal}
  {\bibinfo  {journal} {Phys. Rev. Lett.}\ }\textbf {\bibinfo {volume} {87}},\
  \bibinfo {pages} {157002} (\bibinfo {year} {2001})}\BibitemShut {NoStop}%
\bibitem [{\citenamefont {Choi}\ \emph {et~al.}(2002)\citenamefont {Choi},
  \citenamefont {Roundy}, \citenamefont {Sun}, \citenamefont {Cohen},\ and\
  \citenamefont {Louie}}]{Choi2002}%
  \BibitemOpen
  \bibfield  {author} {\bibinfo {author} {\bibfnamefont {H.~J.}\ \bibnamefont
  {Choi}}, \bibinfo {author} {\bibfnamefont {D.}~\bibnamefont {Roundy}},
  \bibinfo {author} {\bibfnamefont {H.}~\bibnamefont {Sun}}, \bibinfo {author}
  {\bibfnamefont {M.~L.}\ \bibnamefont {Cohen}},\ and\ \bibinfo {author}
  {\bibfnamefont {S.~G.}\ \bibnamefont {Louie}},\ }\bibfield  {title} {\bibinfo
  {title} {The origin of the anomalous superconducting properties of
  {MgB$_2$}},\ }\href {https://doi.org/10.1038/nature00898} {\bibfield
  {journal} {\bibinfo  {journal} {Nature}\ }\textbf {\bibinfo {volume} {418}},\
  \bibinfo {pages} {758} (\bibinfo {year} {2002})}\BibitemShut {NoStop}%
\bibitem [{\citenamefont {Souma}\ \emph {et~al.}(2003)\citenamefont {Souma},
  \citenamefont {Machida}, \citenamefont {Sato}, \citenamefont {Takahashi},
  \citenamefont {Matsui}, \citenamefont {Wang}, \citenamefont {Ding},
  \citenamefont {Kaminski}, \citenamefont {Campuzano}, \citenamefont {Sasaki},\
  and\ \citenamefont {Kadowaki}}]{Souma2003}%
  \BibitemOpen
  \bibfield  {author} {\bibinfo {author} {\bibfnamefont {S.}~\bibnamefont
  {Souma}}, \bibinfo {author} {\bibfnamefont {Y.}~\bibnamefont {Machida}},
  \bibinfo {author} {\bibfnamefont {T.}~\bibnamefont {Sato}}, \bibinfo {author}
  {\bibfnamefont {T.}~\bibnamefont {Takahashi}}, \bibinfo {author}
  {\bibfnamefont {H.}~\bibnamefont {Matsui}}, \bibinfo {author} {\bibfnamefont
  {S.-C.}\ \bibnamefont {Wang}}, \bibinfo {author} {\bibfnamefont
  {H.}~\bibnamefont {Ding}}, \bibinfo {author} {\bibfnamefont {A.}~\bibnamefont
  {Kaminski}}, \bibinfo {author} {\bibfnamefont {J.~C.}\ \bibnamefont
  {Campuzano}}, \bibinfo {author} {\bibfnamefont {S.}~\bibnamefont {Sasaki}},\
  and\ \bibinfo {author} {\bibfnamefont {K.}~\bibnamefont {Kadowaki}},\
  }\bibfield  {title} {\bibinfo {title} {The origin of multiple superconducting
  gaps in {MgB$_2$}},\ }\href {https://doi.org/10.1038/nature01619} {\bibfield
  {journal} {\bibinfo  {journal} {Nature}\ }\textbf {\bibinfo {volume} {423}},\
  \bibinfo {pages} {65} (\bibinfo {year} {2003})}\BibitemShut {NoStop}%
\bibitem [{\citenamefont {Jin}\ \emph {et~al.}(2019)\citenamefont {Jin},
  \citenamefont {Huang}, \citenamefont {Mei}, \citenamefont {Liu},
  \citenamefont {Lim},\ and\ \citenamefont {Liu}}]{Jin:2019}%
  \BibitemOpen
  \bibfield  {author} {\bibinfo {author} {\bibfnamefont {K.-H.}\ \bibnamefont
  {Jin}}, \bibinfo {author} {\bibfnamefont {H.}~\bibnamefont {Huang}}, \bibinfo
  {author} {\bibfnamefont {J.-W.}\ \bibnamefont {Mei}}, \bibinfo {author}
  {\bibfnamefont {Z.}~\bibnamefont {Liu}}, \bibinfo {author} {\bibfnamefont
  {L.-K.}\ \bibnamefont {Lim}},\ and\ \bibinfo {author} {\bibfnamefont
  {F.}~\bibnamefont {Liu}},\ }\bibfield  {title} {\bibinfo {title} {Topological
  superconducting phase in high-tc superconductor {MgB$_2$} with
  dirac--nodal-line fermions},\ }\href
  {https://doi.org/10.1038/s41524-019-0191-2} {\bibfield  {journal} {\bibinfo
  {journal} {npj Computational Materials}\ }\textbf {\bibinfo {volume} {5}},\
  \bibinfo {pages} {57} (\bibinfo {year} {2019})}\BibitemShut {NoStop}%
\bibitem [{\citenamefont {Zhou}\ \emph {et~al.}(2019)\citenamefont {Zhou},
  \citenamefont {Gordon}, \citenamefont {Jin}, \citenamefont {Li},
  \citenamefont {Narayan}, \citenamefont {Zhao}, \citenamefont {Zheng},
  \citenamefont {Huang}, \citenamefont {Cao}, \citenamefont {Zhigadlo},
  \citenamefont {Liu},\ and\ \citenamefont {Dessau}}]{Zhou2019}%
  \BibitemOpen
  \bibfield  {author} {\bibinfo {author} {\bibfnamefont {X.}~\bibnamefont
  {Zhou}}, \bibinfo {author} {\bibfnamefont {K.~N.}\ \bibnamefont {Gordon}},
  \bibinfo {author} {\bibfnamefont {K.-H.}\ \bibnamefont {Jin}}, \bibinfo
  {author} {\bibfnamefont {H.}~\bibnamefont {Li}}, \bibinfo {author}
  {\bibfnamefont {D.}~\bibnamefont {Narayan}}, \bibinfo {author} {\bibfnamefont
  {H.}~\bibnamefont {Zhao}}, \bibinfo {author} {\bibfnamefont {H.}~\bibnamefont
  {Zheng}}, \bibinfo {author} {\bibfnamefont {H.}~\bibnamefont {Huang}},
  \bibinfo {author} {\bibfnamefont {G.}~\bibnamefont {Cao}}, \bibinfo {author}
  {\bibfnamefont {N.~D.}\ \bibnamefont {Zhigadlo}}, \bibinfo {author}
  {\bibfnamefont {F.}~\bibnamefont {Liu}},\ and\ \bibinfo {author}
  {\bibfnamefont {D.~S.}\ \bibnamefont {Dessau}},\ }\bibfield  {title}
  {\bibinfo {title} {Observation of topological surface states in the
  high-temperature superconductor {MgB$_2$}},\ }\href
  {https://doi.org/10.1103/PhysRevB.100.184511} {\bibfield  {journal} {\bibinfo
   {journal} {Phys. Rev. B}\ }\textbf {\bibinfo {volume} {100}},\ \bibinfo
  {pages} {184511} (\bibinfo {year} {2019})}\BibitemShut {NoStop}%
\bibitem [{\citenamefont {An}\ and\ \citenamefont {Pickett}(2001)}]{An:2001}%
  \BibitemOpen
  \bibfield  {author} {\bibinfo {author} {\bibfnamefont {J.~M.}\ \bibnamefont
  {An}}\ and\ \bibinfo {author} {\bibfnamefont {W.~E.}\ \bibnamefont
  {Pickett}},\ }\bibfield  {title} {\bibinfo {title} {Superconductivity of
  {MgB$_2$}: Covalent bonds driven metallic},\ }\href
  {https://doi.org/10.1103/PhysRevLett.86.4366} {\bibfield  {journal} {\bibinfo
   {journal} {Phys. Rev. Lett.}\ }\textbf {\bibinfo {volume} {86}},\ \bibinfo
  {pages} {4366} (\bibinfo {year} {2001})}\BibitemShut {NoStop}%
\bibitem [{\citenamefont {Yu}\ \emph {et~al.}(2024)\citenamefont {Yu},
  \citenamefont {Ciccarino}, \citenamefont {Bianco}, \citenamefont {Errea},
  \citenamefont {Narang},\ and\ \citenamefont {Bernevig}}]{Yu2024}%
  \BibitemOpen
  \bibfield  {author} {\bibinfo {author} {\bibfnamefont {J.}~\bibnamefont
  {Yu}}, \bibinfo {author} {\bibfnamefont {C.~J.}\ \bibnamefont {Ciccarino}},
  \bibinfo {author} {\bibfnamefont {R.}~\bibnamefont {Bianco}}, \bibinfo
  {author} {\bibfnamefont {I.}~\bibnamefont {Errea}}, \bibinfo {author}
  {\bibfnamefont {P.}~\bibnamefont {Narang}},\ and\ \bibinfo {author}
  {\bibfnamefont {B.~A.}\ \bibnamefont {Bernevig}},\ }\bibfield  {title}
  {\bibinfo {title} {Non-trivial quantum geometry and the strength of
  electron–phonon coupling},\ }\href
  {https://doi.org/10.1038/s41567-024-02486-0} {\bibfield  {journal} {\bibinfo
  {journal} {Nature Physics}\ }\textbf {\bibinfo {volume} {20}},\ \bibinfo
  {pages} {1262} (\bibinfo {year} {2024})}\BibitemShut {NoStop}%
\bibitem [{\citenamefont {Mazin}\ and\ \citenamefont
  {Antropov}(2003)}]{Mazin:2003}%
  \BibitemOpen
  \bibfield  {author} {\bibinfo {author} {\bibfnamefont {I.}~\bibnamefont
  {Mazin}}\ and\ \bibinfo {author} {\bibfnamefont {V.}~\bibnamefont
  {Antropov}},\ }\bibfield  {title} {\bibinfo {title} {{Electronic structure,
  electron–phonon coupling, and multiband effects in MgB$_2$}},\ }\href
  {https://doi.org/https://doi.org/10.1016/S0921-4534(02)02299-2} {\bibfield
  {journal} {\bibinfo  {journal} {Physica C: Superconductivity}\ }\textbf
  {\bibinfo {volume} {385}},\ \bibinfo {pages} {49} (\bibinfo {year}
  {2003})}\BibitemShut {NoStop}%
\bibitem [{\citenamefont {Silkin}\ \emph {et~al.}(2009)\citenamefont {Silkin},
  \citenamefont {Balassis}, \citenamefont {Echenique},\ and\ \citenamefont
  {Chulkov}}]{Silkin:2009}%
  \BibitemOpen
  \bibfield  {author} {\bibinfo {author} {\bibfnamefont {V.~M.}\ \bibnamefont
  {Silkin}}, \bibinfo {author} {\bibfnamefont {A.}~\bibnamefont {Balassis}},
  \bibinfo {author} {\bibfnamefont {P.~M.}\ \bibnamefont {Echenique}},\ and\
  \bibinfo {author} {\bibfnamefont {E.~V.}\ \bibnamefont {Chulkov}},\
  }\bibfield  {title} {\bibinfo {title} {Ab initio calculation of low-energy
  collective charge-density excitations in {MgB$_2$}},\ }\href
  {https://doi.org/10.1103/PhysRevB.80.054521} {\bibfield  {journal} {\bibinfo
  {journal} {Phys. Rev. B}\ }\textbf {\bibinfo {volume} {80}},\ \bibinfo
  {pages} {054521} (\bibinfo {year} {2009})}\BibitemShut {NoStop}%
\bibitem [{\citenamefont {Sharma}\ \emph {et~al.}(2011)\citenamefont {Sharma},
  \citenamefont {Kumar}, \citenamefont {Vajpayee}, \citenamefont {Kumar},
  \citenamefont {Ahluwalia},\ and\ \citenamefont {Awana}}]{Sharma2011}%
  \BibitemOpen
  \bibfield  {author} {\bibinfo {author} {\bibfnamefont {D.}~\bibnamefont
  {Sharma}}, \bibinfo {author} {\bibfnamefont {J.}~\bibnamefont {Kumar}},
  \bibinfo {author} {\bibfnamefont {A.}~\bibnamefont {Vajpayee}}, \bibinfo
  {author} {\bibfnamefont {R.}~\bibnamefont {Kumar}}, \bibinfo {author}
  {\bibfnamefont {P.~K.}\ \bibnamefont {Ahluwalia}},\ and\ \bibinfo {author}
  {\bibfnamefont {V.~P.~S.}\ \bibnamefont {Awana}},\ }\bibfield  {title}
  {\bibinfo {title} {Comparative experimental and density functional theory
  ({DFT}) study of the physical properties of {MgB$_2$} and {AlB$_2$}},\ }\href
  {https://doi.org/10.1007/s10948-011-1146-0} {\bibfield  {journal} {\bibinfo
  {journal} {Journal of Superconductivity and Novel Magnetism}\ }\textbf
  {\bibinfo {volume} {24}},\ \bibinfo {pages} {1925} (\bibinfo {year}
  {2011})}\BibitemShut {NoStop}%
\bibitem [{\citenamefont {Vinson}\ \emph {et~al.}(2011)\citenamefont {Vinson},
  \citenamefont {Rehr}, \citenamefont {Kas},\ and\ \citenamefont
  {Shirley}}]{Vinson_BSE}%
  \BibitemOpen
  \bibfield  {author} {\bibinfo {author} {\bibfnamefont {J.}~\bibnamefont
  {Vinson}}, \bibinfo {author} {\bibfnamefont {J.~J.}\ \bibnamefont {Rehr}},
  \bibinfo {author} {\bibfnamefont {J.~J.}\ \bibnamefont {Kas}},\ and\ \bibinfo
  {author} {\bibfnamefont {E.~L.}\ \bibnamefont {Shirley}},\ }\bibfield
  {title} {\bibinfo {title} {Bethe-salpeter equation calculations of core
  excitation spectra},\ }\href {https://doi.org/10.1103/PhysRevB.83.115106}
  {\bibfield  {journal} {\bibinfo  {journal} {Phys. Rev. B}\ }\textbf {\bibinfo
  {volume} {83}},\ \bibinfo {pages} {115106} (\bibinfo {year}
  {2011})}\BibitemShut {NoStop}%
\bibitem [{\citenamefont {Vinson}(2022)}]{Vinson_OCEAN3}%
  \BibitemOpen
  \bibfield  {author} {\bibinfo {author} {\bibfnamefont {J.}~\bibnamefont
  {Vinson}},\ }\bibfield  {title} {\bibinfo {title} {Advances in the {OCEAN-3}
  spectroscopy package},\ }\href {https://doi.org/10.1039/D2CP01030E}
  {\bibfield  {journal} {\bibinfo  {journal} {Phys. Chem. Chem. Phys.}\
  }\textbf {\bibinfo {volume} {24}},\ \bibinfo {pages} {12787} (\bibinfo {year}
  {2022})}\BibitemShut {NoStop}%
\bibitem [{\citenamefont {Giannozzi}\ \emph {et~al.}(2017)\citenamefont
  {Giannozzi}, \citenamefont {Andreussi}, \citenamefont {Brumme}, \citenamefont
  {Bunau}, \citenamefont {Nardelli}, \citenamefont {Calandra}, \citenamefont
  {Car}, \citenamefont {Cavazzoni}, \citenamefont {Ceresoli}, \citenamefont
  {Cococcioni}, \citenamefont {Colonna}, \citenamefont {Carnimeo},
  \citenamefont {Corso}, \citenamefont {de~Gironcoli}, \citenamefont {Delugas},
  \citenamefont {DiStasio}, \citenamefont {Ferretti}, \citenamefont {Floris},
  \citenamefont {Fratesi}, \citenamefont {Fugallo}, \citenamefont {Gebauer},
  \citenamefont {Gerstmann}, \citenamefont {Giustino}, \citenamefont {Gorni},
  \citenamefont {Jia}, \citenamefont {Kawamura}, \citenamefont {Ko},
  \citenamefont {Kokalj}, \citenamefont {Küçükbenli}, \citenamefont
  {Lazzeri}, \citenamefont {Marsili}, \citenamefont {Marzari}, \citenamefont
  {Mauri}, \citenamefont {Nguyen}, \citenamefont {Nguyen}, \citenamefont {de-la
  Roza}, \citenamefont {Paulatto}, \citenamefont {Poncé}, \citenamefont
  {Rocca}, \citenamefont {Sabatini}, \citenamefont {Santra}, \citenamefont
  {Schlipf}, \citenamefont {Seitsonen}, \citenamefont {Smogunov}, \citenamefont
  {Timrov}, \citenamefont {Thonhauser}, \citenamefont {Umari}, \citenamefont
  {Vast}, \citenamefont {Wu},\ and\ \citenamefont
  {Baroni}}]{Giannozzi_QEspresso}%
  \BibitemOpen
  \bibfield  {author} {\bibinfo {author} {\bibfnamefont {P.}~\bibnamefont
  {Giannozzi}}, \bibinfo {author} {\bibfnamefont {O.}~\bibnamefont
  {Andreussi}}, \bibinfo {author} {\bibfnamefont {T.}~\bibnamefont {Brumme}},
  \bibinfo {author} {\bibfnamefont {O.}~\bibnamefont {Bunau}}, \bibinfo
  {author} {\bibfnamefont {M.~B.}\ \bibnamefont {Nardelli}}, \bibinfo {author}
  {\bibfnamefont {M.}~\bibnamefont {Calandra}}, \bibinfo {author}
  {\bibfnamefont {R.}~\bibnamefont {Car}}, \bibinfo {author} {\bibfnamefont
  {C.}~\bibnamefont {Cavazzoni}}, \bibinfo {author} {\bibfnamefont
  {D.}~\bibnamefont {Ceresoli}}, \bibinfo {author} {\bibfnamefont
  {M.}~\bibnamefont {Cococcioni}}, \bibinfo {author} {\bibfnamefont
  {N.}~\bibnamefont {Colonna}}, \bibinfo {author} {\bibfnamefont
  {I.}~\bibnamefont {Carnimeo}}, \bibinfo {author} {\bibfnamefont {A.~D.}\
  \bibnamefont {Corso}}, \bibinfo {author} {\bibfnamefont {S.}~\bibnamefont
  {de~Gironcoli}}, \bibinfo {author} {\bibfnamefont {P.}~\bibnamefont
  {Delugas}}, \bibinfo {author} {\bibfnamefont {R.~A.}\ \bibnamefont
  {DiStasio}}, \bibinfo {author} {\bibfnamefont {A.}~\bibnamefont {Ferretti}},
  \bibinfo {author} {\bibfnamefont {A.}~\bibnamefont {Floris}}, \bibinfo
  {author} {\bibfnamefont {G.}~\bibnamefont {Fratesi}}, \bibinfo {author}
  {\bibfnamefont {G.}~\bibnamefont {Fugallo}}, \bibinfo {author} {\bibfnamefont
  {R.}~\bibnamefont {Gebauer}}, \bibinfo {author} {\bibfnamefont
  {U.}~\bibnamefont {Gerstmann}}, \bibinfo {author} {\bibfnamefont
  {F.}~\bibnamefont {Giustino}}, \bibinfo {author} {\bibfnamefont
  {T.}~\bibnamefont {Gorni}}, \bibinfo {author} {\bibfnamefont
  {J.}~\bibnamefont {Jia}}, \bibinfo {author} {\bibfnamefont {M.}~\bibnamefont
  {Kawamura}}, \bibinfo {author} {\bibfnamefont {H.-Y.}\ \bibnamefont {Ko}},
  \bibinfo {author} {\bibfnamefont {A.}~\bibnamefont {Kokalj}}, \bibinfo
  {author} {\bibfnamefont {E.}~\bibnamefont {Küçükbenli}}, \bibinfo {author}
  {\bibfnamefont {M.}~\bibnamefont {Lazzeri}}, \bibinfo {author} {\bibfnamefont
  {M.}~\bibnamefont {Marsili}}, \bibinfo {author} {\bibfnamefont
  {N.}~\bibnamefont {Marzari}}, \bibinfo {author} {\bibfnamefont
  {F.}~\bibnamefont {Mauri}}, \bibinfo {author} {\bibfnamefont {N.~L.}\
  \bibnamefont {Nguyen}}, \bibinfo {author} {\bibfnamefont {H.-V.}\
  \bibnamefont {Nguyen}}, \bibinfo {author} {\bibfnamefont {A.~O.}\
  \bibnamefont {de-la Roza}}, \bibinfo {author} {\bibfnamefont
  {L.}~\bibnamefont {Paulatto}}, \bibinfo {author} {\bibfnamefont
  {S.}~\bibnamefont {Poncé}}, \bibinfo {author} {\bibfnamefont
  {D.}~\bibnamefont {Rocca}}, \bibinfo {author} {\bibfnamefont
  {R.}~\bibnamefont {Sabatini}}, \bibinfo {author} {\bibfnamefont
  {B.}~\bibnamefont {Santra}}, \bibinfo {author} {\bibfnamefont
  {M.}~\bibnamefont {Schlipf}}, \bibinfo {author} {\bibfnamefont {A.~P.}\
  \bibnamefont {Seitsonen}}, \bibinfo {author} {\bibfnamefont {A.}~\bibnamefont
  {Smogunov}}, \bibinfo {author} {\bibfnamefont {I.}~\bibnamefont {Timrov}},
  \bibinfo {author} {\bibfnamefont {T.}~\bibnamefont {Thonhauser}}, \bibinfo
  {author} {\bibfnamefont {P.}~\bibnamefont {Umari}}, \bibinfo {author}
  {\bibfnamefont {N.}~\bibnamefont {Vast}}, \bibinfo {author} {\bibfnamefont
  {X.}~\bibnamefont {Wu}},\ and\ \bibinfo {author} {\bibfnamefont
  {S.}~\bibnamefont {Baroni}},\ }\bibfield  {title} {\bibinfo {title} {Advanced
  capabilities for materials modelling with quantum espresso},\ }\href
  {https://doi.org/10.1088/1361-648X/aa8f79} {\bibfield  {journal} {\bibinfo
  {journal} {Journal of Physics: Condensed Matter}\ }\textbf {\bibinfo {volume}
  {29}},\ \bibinfo {pages} {465901} (\bibinfo {year} {2017})}\BibitemShut
  {NoStop}%
\bibitem [{\citenamefont {Hamann}(2013)}]{Hamann_ONCVPSP}%
  \BibitemOpen
  \bibfield  {author} {\bibinfo {author} {\bibfnamefont {D.~R.}\ \bibnamefont
  {Hamann}},\ }\bibfield  {title} {\bibinfo {title} {Optimized norm-conserving
  vanderbilt pseudopotentials},\ }\href
  {https://doi.org/10.1103/PhysRevB.88.085117} {\bibfield  {journal} {\bibinfo
  {journal} {Phys. Rev. B}\ }\textbf {\bibinfo {volume} {88}},\ \bibinfo
  {pages} {085117} (\bibinfo {year} {2013})}\BibitemShut {NoStop}%
\bibitem [{\citenamefont {Sun}\ \emph {et~al.}(2015)\citenamefont {Sun},
  \citenamefont {Ruzsinszky},\ and\ \citenamefont {Perdew}}]{Sun_SCAN}%
  \BibitemOpen
  \bibfield  {author} {\bibinfo {author} {\bibfnamefont {J.}~\bibnamefont
  {Sun}}, \bibinfo {author} {\bibfnamefont {A.}~\bibnamefont {Ruzsinszky}},\
  and\ \bibinfo {author} {\bibfnamefont {J.~P.}\ \bibnamefont {Perdew}},\
  }\bibfield  {title} {\bibinfo {title} {Strongly constrained and appropriately
  normed semilocal density functional},\ }\href
  {https://doi.org/10.1103/PhysRevLett.115.036402} {\bibfield  {journal}
  {\bibinfo  {journal} {Phys. Rev. Lett.}\ }\textbf {\bibinfo {volume} {115}},\
  \bibinfo {pages} {036402} (\bibinfo {year} {2015})}\BibitemShut {NoStop}%
\bibitem [{\citenamefont {Zhu}\ \emph {et~al.}(2002)\citenamefont {Zhu},
  \citenamefont {Moodenbaugh}, \citenamefont {Schneider}, \citenamefont
  {Davenport}, \citenamefont {Vogt}, \citenamefont {Li}, \citenamefont {Gu},
  \citenamefont {Fischer},\ and\ \citenamefont {Tafto}}]{Zhu:2002}%
  \BibitemOpen
  \bibfield  {author} {\bibinfo {author} {\bibfnamefont {Y.}~\bibnamefont
  {Zhu}}, \bibinfo {author} {\bibfnamefont {A.~R.}\ \bibnamefont
  {Moodenbaugh}}, \bibinfo {author} {\bibfnamefont {G.}~\bibnamefont
  {Schneider}}, \bibinfo {author} {\bibfnamefont {J.~W.}\ \bibnamefont
  {Davenport}}, \bibinfo {author} {\bibfnamefont {T.}~\bibnamefont {Vogt}},
  \bibinfo {author} {\bibfnamefont {Q.}~\bibnamefont {Li}}, \bibinfo {author}
  {\bibfnamefont {G.}~\bibnamefont {Gu}}, \bibinfo {author} {\bibfnamefont
  {D.~A.}\ \bibnamefont {Fischer}},\ and\ \bibinfo {author} {\bibfnamefont
  {J.}~\bibnamefont {Tafto}},\ }\bibfield  {title} {\bibinfo {title}
  {Unraveling the symmetry of the hole states near the fermi level in the
  {MgB$_2$} superconductor},\ }\href
  {https://doi.org/10.1103/PhysRevLett.88.247002} {\bibfield  {journal}
  {\bibinfo  {journal} {Phys. Rev. Lett.}\ }\textbf {\bibinfo {volume} {88}},\
  \bibinfo {pages} {247002} (\bibinfo {year} {2002})}\BibitemShut {NoStop}%
\bibitem [{\citenamefont {Saitoh}\ \emph {et~al.}(2012)\citenamefont {Saitoh},
  \citenamefont {Momonoi}, \citenamefont {Tanaka},\ and\ \citenamefont
  {Onari}}]{Saito:2012}%
  \BibitemOpen
  \bibfield  {author} {\bibinfo {author} {\bibfnamefont {K.}~\bibnamefont
  {Saitoh}}, \bibinfo {author} {\bibfnamefont {K.}~\bibnamefont {Momonoi}},
  \bibinfo {author} {\bibfnamefont {N.}~\bibnamefont {Tanaka}},\ and\ \bibinfo
  {author} {\bibfnamefont {S.}~\bibnamefont {Onari}},\ }\bibfield  {title}
  {\bibinfo {title} {Observation of the hole state symmetry of {MgB$_2$} by
  inelastic scattering of fast electrons accompanied by boron k-shell
  excitation},\ }\href {https://doi.org/10.1063/1.4768728} {\bibfield
  {journal} {\bibinfo  {journal} {Journal of Applied Physics}\ }\textbf
  {\bibinfo {volume} {112}},\ \bibinfo {pages} {113920} (\bibinfo {year}
  {2012})}\BibitemShut {NoStop}%
\bibitem [{\citenamefont {Idrobo}\ and\ \citenamefont
  {Browning}(2004)}]{Idrobo:2004}%
  \BibitemOpen
  \bibfield  {author} {\bibinfo {author} {\bibfnamefont {J.~C.}\ \bibnamefont
  {Idrobo}}\ and\ \bibinfo {author} {\bibfnamefont {N.~D.}\ \bibnamefont
  {Browning}},\ }\bibfield  {title} {\bibinfo {title} {Distinguishing
  intra-band and inter-band transitions in {MgB$_2$} using monochromated
  eels},\ }\href {https://doi.org/10.1017/S1431927604885283} {\bibfield
  {journal} {\bibinfo  {journal} {Microscopy and Microanalysis}\ }\textbf
  {\bibinfo {volume} {10}},\ \bibinfo {pages} {840} (\bibinfo {year}
  {2004})}\BibitemShut {NoStop}%
\bibitem [{\citenamefont {Zhukov}\ \emph {et~al.}(2001)\citenamefont {Zhukov},
  \citenamefont {Silkin}, \citenamefont {Chulkov},\ and\ \citenamefont
  {Echenique}}]{Zhukov:2001}%
  \BibitemOpen
  \bibfield  {author} {\bibinfo {author} {\bibfnamefont {V.~P.}\ \bibnamefont
  {Zhukov}}, \bibinfo {author} {\bibfnamefont {V.~M.}\ \bibnamefont {Silkin}},
  \bibinfo {author} {\bibfnamefont {E.~V.}\ \bibnamefont {Chulkov}},\ and\
  \bibinfo {author} {\bibfnamefont {P.~M.}\ \bibnamefont {Echenique}},\
  }\bibfield  {title} {\bibinfo {title} {Dielectric functions and collective
  excitations in {MgB$_2$}},\ }\href
  {https://doi.org/10.1103/PhysRevB.64.180507} {\bibfield  {journal} {\bibinfo
  {journal} {Phys. Rev. B}\ }\textbf {\bibinfo {volume} {64}},\ \bibinfo
  {pages} {180507} (\bibinfo {year} {2001})}\BibitemShut {NoStop}%
\bibitem [{\citenamefont {Mitrano}\ \emph {et~al.}(2018)\citenamefont
  {Mitrano}, \citenamefont {Husain}, \citenamefont {Vig}, \citenamefont
  {Kogar}, \citenamefont {Rak}, \citenamefont {Rubeck}, \citenamefont
  {Schmalian}, \citenamefont {Uchoa}, \citenamefont {Schneeloch}, \citenamefont
  {Zhong}, \citenamefont {Gu},\ and\ \citenamefont {Abbamonte}}]{Mitrano:2018}%
  \BibitemOpen
  \bibfield  {author} {\bibinfo {author} {\bibfnamefont {M.}~\bibnamefont
  {Mitrano}}, \bibinfo {author} {\bibfnamefont {A.~A.}\ \bibnamefont {Husain}},
  \bibinfo {author} {\bibfnamefont {S.}~\bibnamefont {Vig}}, \bibinfo {author}
  {\bibfnamefont {A.}~\bibnamefont {Kogar}}, \bibinfo {author} {\bibfnamefont
  {M.~S.}\ \bibnamefont {Rak}}, \bibinfo {author} {\bibfnamefont {S.~I.}\
  \bibnamefont {Rubeck}}, \bibinfo {author} {\bibfnamefont {J.}~\bibnamefont
  {Schmalian}}, \bibinfo {author} {\bibfnamefont {B.}~\bibnamefont {Uchoa}},
  \bibinfo {author} {\bibfnamefont {J.}~\bibnamefont {Schneeloch}}, \bibinfo
  {author} {\bibfnamefont {R.}~\bibnamefont {Zhong}}, \bibinfo {author}
  {\bibfnamefont {G.~D.}\ \bibnamefont {Gu}},\ and\ \bibinfo {author}
  {\bibfnamefont {P.}~\bibnamefont {Abbamonte}},\ }\bibfield  {title} {\bibinfo
  {title} {Anomalous density fluctuations in a strange metal},\ }\href
  {https://doi.org/10.1073/pnas.1721495115} {\bibfield  {journal} {\bibinfo
  {journal} {Proceedings of the National Academy of Sciences}\ }\textbf
  {\bibinfo {volume} {115}},\ \bibinfo {pages} {5392} (\bibinfo {year}
  {2018})}\BibitemShut {NoStop}%
\bibitem [{\citenamefont {Xi}\ \emph {et~al.}(2007)\citenamefont {Xi},
  \citenamefont {Pogrebnyakov}, \citenamefont {Xu}, \citenamefont {Chen},
  \citenamefont {Cui}, \citenamefont {Maertz}, \citenamefont {Zhuang},
  \citenamefont {Li}, \citenamefont {Lamborn}, \citenamefont {Redwing},
  \citenamefont {Liu}, \citenamefont {Soukiassian}, \citenamefont {Schlom},
  \citenamefont {Weng}, \citenamefont {Dickey}, \citenamefont {Chen},
  \citenamefont {Tian}, \citenamefont {Pan}, \citenamefont {Cybart},\ and\
  \citenamefont {Dynes}}]{Xi:2007}%
  \BibitemOpen
  \bibfield  {author} {\bibinfo {author} {\bibfnamefont {X.}~\bibnamefont
  {Xi}}, \bibinfo {author} {\bibfnamefont {A.}~\bibnamefont {Pogrebnyakov}},
  \bibinfo {author} {\bibfnamefont {S.}~\bibnamefont {Xu}}, \bibinfo {author}
  {\bibfnamefont {K.}~\bibnamefont {Chen}}, \bibinfo {author} {\bibfnamefont
  {Y.}~\bibnamefont {Cui}}, \bibinfo {author} {\bibfnamefont {E.}~\bibnamefont
  {Maertz}}, \bibinfo {author} {\bibfnamefont {C.}~\bibnamefont {Zhuang}},
  \bibinfo {author} {\bibfnamefont {Q.}~\bibnamefont {Li}}, \bibinfo {author}
  {\bibfnamefont {D.}~\bibnamefont {Lamborn}}, \bibinfo {author} {\bibfnamefont
  {J.}~\bibnamefont {Redwing}}, \bibinfo {author} {\bibfnamefont
  {Z.}~\bibnamefont {Liu}}, \bibinfo {author} {\bibfnamefont {A.}~\bibnamefont
  {Soukiassian}}, \bibinfo {author} {\bibfnamefont {D.}~\bibnamefont {Schlom}},
  \bibinfo {author} {\bibfnamefont {X.}~\bibnamefont {Weng}}, \bibinfo {author}
  {\bibfnamefont {E.}~\bibnamefont {Dickey}}, \bibinfo {author} {\bibfnamefont
  {Y.}~\bibnamefont {Chen}}, \bibinfo {author} {\bibfnamefont {W.}~\bibnamefont
  {Tian}}, \bibinfo {author} {\bibfnamefont {X.}~\bibnamefont {Pan}}, \bibinfo
  {author} {\bibfnamefont {S.}~\bibnamefont {Cybart}},\ and\ \bibinfo {author}
  {\bibfnamefont {R.}~\bibnamefont {Dynes}},\ }\bibfield  {title} {\bibinfo
  {title} {{MgB$_2$ thin films by hybrid physical–chemical vapor
  deposition}},\ }\href
  {https://doi.org/https://doi.org/10.1016/j.physc.2007.01.029} {\bibfield
  {journal} {\bibinfo  {journal} {Physica C: Superconductivity}\ }\textbf
  {\bibinfo {volume} {456}},\ \bibinfo {pages} {22} (\bibinfo {year} {2007})},\
  \bibinfo {note} {recent Advances in MgB2 Research}\BibitemShut {NoStop}%
\bibitem [{\citenamefont {{van Setten}}\ \emph {et~al.}(2018)\citenamefont
  {{van Setten}}, \citenamefont {Giantomassi}, \citenamefont {Bousquet},
  \citenamefont {Verstraete}, \citenamefont {Hamann}, \citenamefont {Gonze},\
  and\ \citenamefont {Rignanese}}]{vanSetten:2018}%
  \BibitemOpen
  \bibfield  {author} {\bibinfo {author} {\bibfnamefont {M.}~\bibnamefont {{van
  Setten}}}, \bibinfo {author} {\bibfnamefont {M.}~\bibnamefont {Giantomassi}},
  \bibinfo {author} {\bibfnamefont {E.}~\bibnamefont {Bousquet}}, \bibinfo
  {author} {\bibfnamefont {M.}~\bibnamefont {Verstraete}}, \bibinfo {author}
  {\bibfnamefont {D.}~\bibnamefont {Hamann}}, \bibinfo {author} {\bibfnamefont
  {X.}~\bibnamefont {Gonze}},\ and\ \bibinfo {author} {\bibfnamefont {G.-M.}\
  \bibnamefont {Rignanese}},\ }\bibfield  {title} {\bibinfo {title} {The
  pseudodojo: Training and grading a 85 element optimized norm-conserving
  pseudopotential table},\ }\href
  {https://doi.org/https://doi.org/10.1016/j.cpc.2018.01.012} {\bibfield
  {journal} {\bibinfo  {journal} {Computer Physics Communications}\ }\textbf
  {\bibinfo {volume} {226}},\ \bibinfo {pages} {39} (\bibinfo {year}
  {2018})}\BibitemShut {NoStop}%
\bibitem [{\citenamefont {Perdew}\ and\ \citenamefont
  {Wang}(1992)}]{Perdew_LDA}%
  \BibitemOpen
  \bibfield  {author} {\bibinfo {author} {\bibfnamefont {J.~P.}\ \bibnamefont
  {Perdew}}\ and\ \bibinfo {author} {\bibfnamefont {Y.}~\bibnamefont {Wang}},\
  }\bibfield  {title} {\bibinfo {title} {Accurate and simple analytic
  representation of the electron-gas correlation energy},\ }\href
  {https://doi.org/10.1103/PhysRevB.45.13244} {\bibfield  {journal} {\bibinfo
  {journal} {Phys. Rev. B}\ }\textbf {\bibinfo {volume} {45}},\ \bibinfo
  {pages} {13244} (\bibinfo {year} {1992})}\BibitemShut {NoStop}%
\bibitem [{\citenamefont {Perdew}\ \emph {et~al.}(1996)\citenamefont {Perdew},
  \citenamefont {Burke},\ and\ \citenamefont {Ernzerhof}}]{Perdew_PBE}%
  \BibitemOpen
  \bibfield  {author} {\bibinfo {author} {\bibfnamefont {J.~P.}\ \bibnamefont
  {Perdew}}, \bibinfo {author} {\bibfnamefont {K.}~\bibnamefont {Burke}},\ and\
  \bibinfo {author} {\bibfnamefont {M.}~\bibnamefont {Ernzerhof}},\ }\bibfield
  {title} {\bibinfo {title} {Generalized gradient approximation made simple},\
  }\href {https://doi.org/10.1103/PhysRevLett.77.3865} {\bibfield  {journal}
  {\bibinfo  {journal} {Phys. Rev. Lett.}\ }\textbf {\bibinfo {volume} {77}},\
  \bibinfo {pages} {3865} (\bibinfo {year} {1996})}\BibitemShut {NoStop}%
\bibitem [{\citenamefont {Momma}\ and\ \citenamefont
  {Izumi}(2011)}]{Momma_Vesta}%
  \BibitemOpen
  \bibfield  {author} {\bibinfo {author} {\bibfnamefont {K.}~\bibnamefont
  {Momma}}\ and\ \bibinfo {author} {\bibfnamefont {F.}~\bibnamefont {Izumi}},\
  }\bibfield  {title} {\bibinfo {title} {{{\it VESTA3} for three-dimensional
  visualization of crystal, volumetric and morphology data}},\ }\href
  {https://doi.org/10.1107/S0021889811038970} {\bibfield  {journal} {\bibinfo
  {journal} {Journal of Applied Crystallography}\ }\textbf {\bibinfo {volume}
  {44}},\ \bibinfo {pages} {1272} (\bibinfo {year} {2011})}\BibitemShut
  {NoStop}%
\bibitem [{\citenamefont {Kawamura}(2019)}]{Kawamura_FermiSurfer}%
  \BibitemOpen
  \bibfield  {author} {\bibinfo {author} {\bibfnamefont {M.}~\bibnamefont
  {Kawamura}},\ }\bibfield  {title} {\bibinfo {title} {Fermisurfer:
  Fermi-surface viewer providing multiple representation schemes},\ }\href
  {https://doi.org/https://doi.org/10.1016/j.cpc.2019.01.017} {\bibfield
  {journal} {\bibinfo  {journal} {Computer Physics Communications}\ }\textbf
  {\bibinfo {volume} {239}},\ \bibinfo {pages} {197} (\bibinfo {year}
  {2019})}\BibitemShut {NoStop}%
\bibitem [{\citenamefont {Saad}\ and\ \citenamefont {Schultz}(1986)}]{Saad}%
  \BibitemOpen
  \bibfield  {author} {\bibinfo {author} {\bibfnamefont {Y.}~\bibnamefont
  {Saad}}\ and\ \bibinfo {author} {\bibfnamefont {M.~H.}\ \bibnamefont
  {Schultz}},\ }\bibfield  {title} {\bibinfo {title} {Gmres: A generalized
  minimal residual algorithm for solving nonsymmetric linear systems},\ }\href
  {https://doi.org/10.1137/0907058} {\bibfield  {journal} {\bibinfo  {journal}
  {SIAM Journal on Scientific and Statistical Computing}\ }\textbf {\bibinfo
  {volume} {7}},\ \bibinfo {pages} {856} (\bibinfo {year} {1986})}\BibitemShut
  {NoStop}%
\bibitem [{\citenamefont {Mortensen}\ \emph {et~al.}(2024)\citenamefont
  {Mortensen}, \citenamefont {Larsen}, \citenamefont {Kuisma}, \citenamefont
  {Ivanov}, \citenamefont {Taghizadeh}, \citenamefont {Peterson}, \citenamefont
  {Haldar}, \citenamefont {Dohn}, \citenamefont {Sch{\"a}fer}, \citenamefont
  {J\'{o}nsson}, \citenamefont {Hermes}, \citenamefont {Nilsson}, \citenamefont
  {Kastlunger}, \citenamefont {Levi}, \citenamefont {J\'{o}nsson},
  \citenamefont {H{\"a}kkinen}, \citenamefont {Fojt}, \citenamefont
  {Kangsabanik}, \citenamefont {S{\o}dequist}, \citenamefont {Lehtom{\"a}ki},
  \citenamefont {Heske}, \citenamefont {Enkovaara}, \citenamefont {Winther},
  \citenamefont {Dulak}, \citenamefont {Melander}, \citenamefont {Ovesen},
  \citenamefont {Louhivuori}, \citenamefont {Walter}, \citenamefont {Gjerding},
  \citenamefont {Lopez-Acevedo}, \citenamefont {Erhart}, \citenamefont
  {Warmbier}, \citenamefont {W{\"u}rdemann}, \citenamefont {Kaappa},
  \citenamefont {Latini}, \citenamefont {Boland}, \citenamefont {Bligaard},
  \citenamefont {Skovhus}, \citenamefont {Susi}, \citenamefont {Maxson},
  \citenamefont {Rossi}, \citenamefont {Chen}, \citenamefont {Schmerwitz},
  \citenamefont {Schi{\o}tz}, \citenamefont {Olsen}, \citenamefont {Jacobsen},\
  and\ \citenamefont {Thygesen}}]{mortensen_gpaw_2024}%
  \BibitemOpen
  \bibfield  {author} {\bibinfo {author} {\bibfnamefont {J.~J.}\ \bibnamefont
  {Mortensen}}, \bibinfo {author} {\bibfnamefont {A.~H.}\ \bibnamefont
  {Larsen}}, \bibinfo {author} {\bibfnamefont {M.}~\bibnamefont {Kuisma}},
  \bibinfo {author} {\bibfnamefont {A.~V.}\ \bibnamefont {Ivanov}}, \bibinfo
  {author} {\bibfnamefont {A.}~\bibnamefont {Taghizadeh}}, \bibinfo {author}
  {\bibfnamefont {A.}~\bibnamefont {Peterson}}, \bibinfo {author}
  {\bibfnamefont {A.}~\bibnamefont {Haldar}}, \bibinfo {author} {\bibfnamefont
  {A.~O.}\ \bibnamefont {Dohn}}, \bibinfo {author} {\bibfnamefont
  {C.}~\bibnamefont {Sch{\"a}fer}}, \bibinfo {author} {\bibfnamefont {E.~O.}\
  \bibnamefont {J\'{o}nsson}}, \bibinfo {author} {\bibfnamefont {E.~D.}\
  \bibnamefont {Hermes}}, \bibinfo {author} {\bibfnamefont {F.~A.}\
  \bibnamefont {Nilsson}}, \bibinfo {author} {\bibfnamefont {G.}~\bibnamefont
  {Kastlunger}}, \bibinfo {author} {\bibfnamefont {G.}~\bibnamefont {Levi}},
  \bibinfo {author} {\bibfnamefont {H.}~\bibnamefont {J\'{o}nsson}}, \bibinfo
  {author} {\bibfnamefont {H.}~\bibnamefont {H{\"a}kkinen}}, \bibinfo {author}
  {\bibfnamefont {J.}~\bibnamefont {Fojt}}, \bibinfo {author} {\bibfnamefont
  {J.}~\bibnamefont {Kangsabanik}}, \bibinfo {author} {\bibfnamefont
  {J.}~\bibnamefont {S{\o}dequist}}, \bibinfo {author} {\bibfnamefont
  {J.}~\bibnamefont {Lehtom{\"a}ki}}, \bibinfo {author} {\bibfnamefont
  {J.}~\bibnamefont {Heske}}, \bibinfo {author} {\bibfnamefont
  {J.}~\bibnamefont {Enkovaara}}, \bibinfo {author} {\bibfnamefont {K.~T.}\
  \bibnamefont {Winther}}, \bibinfo {author} {\bibfnamefont {M.}~\bibnamefont
  {Dulak}}, \bibinfo {author} {\bibfnamefont {M.~M.}\ \bibnamefont {Melander}},
  \bibinfo {author} {\bibfnamefont {M.}~\bibnamefont {Ovesen}}, \bibinfo
  {author} {\bibfnamefont {M.}~\bibnamefont {Louhivuori}}, \bibinfo {author}
  {\bibfnamefont {M.}~\bibnamefont {Walter}}, \bibinfo {author} {\bibfnamefont
  {M.}~\bibnamefont {Gjerding}}, \bibinfo {author} {\bibfnamefont
  {O.}~\bibnamefont {Lopez-Acevedo}}, \bibinfo {author} {\bibfnamefont
  {P.}~\bibnamefont {Erhart}}, \bibinfo {author} {\bibfnamefont
  {R.}~\bibnamefont {Warmbier}}, \bibinfo {author} {\bibfnamefont
  {R.}~\bibnamefont {W{\"u}rdemann}}, \bibinfo {author} {\bibfnamefont
  {S.}~\bibnamefont {Kaappa}}, \bibinfo {author} {\bibfnamefont
  {S.}~\bibnamefont {Latini}}, \bibinfo {author} {\bibfnamefont {T.~M.}\
  \bibnamefont {Boland}}, \bibinfo {author} {\bibfnamefont {T.}~\bibnamefont
  {Bligaard}}, \bibinfo {author} {\bibfnamefont {T.}~\bibnamefont {Skovhus}},
  \bibinfo {author} {\bibfnamefont {T.}~\bibnamefont {Susi}}, \bibinfo {author}
  {\bibfnamefont {T.}~\bibnamefont {Maxson}}, \bibinfo {author} {\bibfnamefont
  {T.}~\bibnamefont {Rossi}}, \bibinfo {author} {\bibfnamefont
  {X.}~\bibnamefont {Chen}}, \bibinfo {author} {\bibfnamefont {Y.~L.~A.}\
  \bibnamefont {Schmerwitz}}, \bibinfo {author} {\bibfnamefont
  {J.}~\bibnamefont {Schi{\o}tz}}, \bibinfo {author} {\bibfnamefont
  {T.}~\bibnamefont {Olsen}}, \bibinfo {author} {\bibfnamefont {K.~W.}\
  \bibnamefont {Jacobsen}},\ and\ \bibinfo {author} {\bibfnamefont {K.~S.}\
  \bibnamefont {Thygesen}},\ }\bibfield  {title} {\bibinfo {title} {{GPAW: An
  open Python package for electronic structure calculations}},\ }\href
  {https://doi.org/10.1063/5.0182685} {\bibfield  {journal} {\bibinfo
  {journal} {The Journal of Chemical Physics}\ }\textbf {\bibinfo {volume}
  {160}},\ \bibinfo {pages} {092503} (\bibinfo {year} {2024})}\BibitemShut
  {NoStop}%
\end{thebibliography}%
% Produces the bibliography via BibTeX.

\clearpage
\newpage 
\onecolumngrid

\setcounter{page}{1}
\setcounter{subsection}{0}

\renewcommand\thefigure{S\arabic{figure}}
\renewcommand\thepage{S\arabic{page}}
\renewcommand\thesection{S\arabic{section}}
\setcounter{figure}{0}
%\linespread{2}

\section*{Supplementary Material}
\subsection{X-ray scattering}
\label{sec:RIXS}
X-Ray Absorption Spectroscopy (XAS) and Resonant Inelastic X-ray Scattering (RIXS) measurements were performed at the qRIXS beamline of the Advanced Light Source (ALS). The combined energy resolution (beamline and spectrometer) was $\approx 100\,\mathrm{meV}$ at the B $K$-edge (190~eV). The incident photon energy was tuned to optimize the sharp feature in the inelastic spectrum at $\omega <1\,\mathrm{eV}$. The incident angle $\theta_\mathrm{in}$ was changed to measure different points in momentum space. The momentum-dependent measurements were conducted along the $\Gamma-M$ and $\Gamma - K$ directions with units given in reciprocal lattice units (r.l.u.) of the hexagonal unit cell with lattice parameters $a = b = 3.086$~\AA ~and $c = 3.524$~\AA. The sample temperature was set to 77\,K.

\subsection{Sample Growth}
\label{sec:sample}
Sample preparation is detailed in Ref. \cite{Xi:2007}. 

\subsection{OCEAN BSE and Quantum ESPRESSO DFT Calculations}
The {\sc ocean} (v3.2.1) code based on the Bethe-Salpeter equation (BSE) formalism for many-body spectroscopic effects was used for calculating the XAS and RIXS of the Boron $K$-edge of MgB$_2$~\cite{Vinson_BSE,Vinson_OCEAN3}. The wavefunction of our system was generated using the Density Functional Theory (DFT) Quantum {\sc espresso} (v7.4) code, using pseudopotentials from the Optimized Norm-Conserving Vanderbilt PSeudopotential ({\sc oncvpsp}) code,
 scalar-relativistic version 3.3.1~\cite{Giannozzi_QEspresso,Hamann_ONCVPSP,vanSetten:2018}. The meta-generalized gradient approximation (meta-GGA), implemented Strongly Constrained and Appropriately Normed (SCAN), was employed as the exchange-correlation functional~\cite{Sun_SCAN}. The local density approximation (LDA) and the regular generalized gradient approximation (GGA) were also considered, but SCAN reproduced the most consistent XAS profile compared to the experiment [see \solofigureref{fig:XAS_xc_comp}~]~\cite{Perdew_LDA,Perdew_PBE}. Fermi-Dirac smearing with a width of 0.02 Ry was used, with a plane wave cutoff of 85 Ry and a uniform $k$-grid of 36 x 36 x 28. A convergence threshold of $1e-10$ Ry was used. 12 total bands were used, and the first 4 occupied bands (Mg 2{\it s} and 2{\it p}) were neglected for the RIXS calculation. The polarization and momentum dependence followed the same experimental setup. A global energy shift was used to align the calculated spectra to the sharp experimental feature near 187 eV, at the pre-edge. A broadening value of 0.1 eV was used. The crystal structures and Fermi surfaces in \figureref{fig:Fig1_intro}{b,c} were generated using {\sc vesta} and {\sc FermiSurfer}, respectively~\cite{Momma_Vesta,Kawamura_FermiSurfer}. 

 The band structure projections in \figref{fig:Fig2}{c} were computed from the exciton vector $x(\omega)$, as defined in Ref.~\cite{Vinson_OCEAN3}, where the X-ray absorption exciton can be written as
\begin{equation}
\left| x_{\mathbf{q},\hat{\mathcal{E}}}(\omega) \right\rangle = 
\frac{1}{\omega - H_{\mathrm{BSE}} + i\eta} \, T_{\mathbf{q},\hat{\mathcal{E}}} \left| \Phi_0 \right\rangle
\end{equation}
 Where $\textbf{q}$ is the photon momentum transfer, $\hat{\mathcal{E}}$ is the photon polarization vector, $\omega$ is the excitation energy, $ H_{\mathrm{BSE}}$ is the BSE Hamiltonian, $\eta$ is the lifetime broadening, $T$ is the photon operator, and $\left| \Phi_0 \right\rangle$ is the DFT ground state. This can be obtained by solving for $x(\omega)$ using the generalized minimal residual (GMRES) method \cite{Saad}. A similar expression can be given for the RIXS exciton. Within {\sc ocean}, $\vert x(\omega)\rangle$ is a vector indexed by conduction band $c$, k-point $\mathbf{k}$, and spin $\sigma$ as well as core level azimuthal quantum number $m_l$ and spin $s$. With a K-edge excitation (only $m_l=0$) and spin-degenerate conduction bands, we construct the exciton density
 \begin{equation}
     \rho_{c\mathbf{k}} = \sum_{\sigma s}\vert x_{c\mathbf{k}\sigma s}(\omega)\vert^2.
 \end{equation}
 We project this density from the dense regular k-point mesh of the BSE calculation onto the band structure using trilinear interpolation.

\subsection{Tight-binding model for MgB$_{2}$\label{tbm}}

To study the low-energy physics in MgB$_{2}$, one can neglect spin-orbit coupling  for light atoms B and Mg and assume spin SU(2) symmetry, thereby using a spinless model. MgB$_{2}$ has two B atoms and one Mg atom per unit cell. Following Ref.\cite{Yu2024} we only consider the
$s,p_{x},p_{y}$ and $p_{z}$ orbitals at each B atom and an $s$
orbital at each Mg atom, and use the Hamiltonian 
\begin{equation}
H=H^{B,s\hspace{0.5mm}p_{x}\hspace{0.5mm}p_{y}}+H^{B,p_{z}}+H^{Mg}+H^{B,s\hspace{0.5mm}p_{x}\hspace{0.5mm}p_{y}-p_{z}}+H^{Mg-B}.\label{eq:1}
\end{equation}
 Now, for states near the Fermi energy, $H^{B,s\hspace{0.5mm}p_{x}\hspace{0.5mm}p_{y}-p_{z}}$
can be safely neglected, since the hopping $t_{B,s-p_{z},z}$ mainly affects the
bands far from the Fermi energy. Moreover, the electron orbitals near the
Fermi energy are mainly boron orbitals, allowing us to neglect $H^{Mg}$ and $H^{Mg-B}$, and the following
simplified Hamiltonian is considered for our analysis:
\begin{equation}
H=H^{B,s\hspace{0.5mm}p_{x}\hspace{0.5mm}p_{y}}+H^{B,p_{z}}\label{eq:14}
\end{equation}
Here,  
\begin{equation}
H^{B,s\hspace{0.5mm}p_{x}\hspace{0.5mm}p_{y}}=\sum_{\mathbf{k}}^{1BZ}c_{\mathbf{k},B,s\hspace{0.5mm}p_{x}\hspace{0.5mm}p_{y}}^{\dagger}h_{s\hspace{0.5mm}p_{x}\hspace{0.5mm}p_{y}}(\mathbf{k})c_{\mathbf{k},B,s\hspace{0.5mm}p_{x}\hspace{0.5mm}p_{y}}\label{eq:2}
\end{equation}
 with $c_{\mathbf{k},B,s\hspace{0.5mm}p_{x}\hspace{0.5mm}p_{y}}^{\dagger}=\left(\begin{array}{cccccc}
c_{\mathbf{k},\gamma_{B_{1}},s}^{\dagger} & c_{\mathbf{k},\gamma_{B_{1}},p_{x}}^{\dagger} & c_{\mathbf{k},\gamma_{B_{1}},p_{y}}^{\dagger} & c_{\mathbf{k},\gamma_{B_{2}},s}^{\dagger} & c_{\mathbf{k},\gamma_{B_{2}},p_{x}}^{\dagger} & c_{\mathbf{k},\gamma_{B_{2}},p_{y}}^{\dagger}\end{array}\right)$, 
\begin{equation}
h_{sp_{x}p_{y}}(\mathbf{k})=\left(\begin{array}{cc}
h_{B1-B1,s\hspace{0.5mm}p_{x}\hspace{0.5mm}p_{y}}(\mathbf{k}) & h_{B1-B2,s\hspace{0.5mm}p_{x}\hspace{0.5mm}p_{y}}(\mathbf{k})\\
h_{B2-B1,s\hspace{0.5mm}p_{x}\hspace{0.5mm}p_{y}}(\mathbf{k}) & h_{B2-B2,s\hspace{0.5mm}p_{x}\hspace{0.5mm}p_{y}}(\mathbf{k})
\end{array}\right)\label{eq:3}
\end{equation}
\begin{equation}
h_{B1-B1,s\hspace{0.5mm}p_{x}\hspace{0.5mm}p_{y}}(\mathbf{k})=\left(\begin{array}{ccc}
E_{B,s,0}\\
 & E_{B,p_{x}\hspace{0.5mm}p_{y},0}\\
 &  & E_{B,p_{x}\hspace{0.5mm}p_{y},0}
\end{array}\right)+\left(\begin{array}{ccc}
t_{B,s,z}2\cos(\mathbf{a_{3}}.\mathbf{k})\\
 & t_{B,p_{x}\hspace{0.5mm}p_{y},z}2\cos(\mathbf{a_{3}}.\mathbf{k})\\
 &  & t_{B,p_{x}\hspace{0.5mm}p_{y},z}2\cos(\mathbf{a_{3}}.\mathbf{k})
\end{array}\right)\label{eq:4}
\end{equation}
\begin{equation}
h_{B2-B2,s\hspace{0.5mm}p_{x}\hspace{0.5mm}p_{y}}(\mathbf{k})=\left(\begin{array}{ccc}
1\\
 & -1\\
 &  & -1
\end{array}\right)h_{B1-B1,s\hspace{0.5mm}p_{x}\hspace{0.5mm}p_{y}}(-\mathbf{k})\left(\begin{array}{ccc}
1\\
 & -1\\
 &  & -1
\end{array}\right)\label{eq:5}
\end{equation}
\begin{equation}
h_{B1-B2,s\hspace{0.5mm}p_{x}\hspace{0.5mm}p_{y}}(\mathbf{k})=h_{B2-B1,s\hspace{0.5mm}p_{x}\hspace{0.5mm}p_{y}}^{\dagger}(\mathbf{k})=\sum_{j=0,1,2}\exp\{-i\mathbf{\delta_{j}}.\mathbf{k}\}\left(\begin{array}{ccc}
1\\
 & -\frac{1}{2} & -\frac{\sqrt{3}}{2}\\
 & \frac{\sqrt{3}}{2} & -\frac{1}{2}
\end{array}\right)^{j}\left(\begin{array}{ccc}
t_{1} &  & t_{4}\\
 & t_{2}+t_{3}\\
-t_{4} &  & t_{2}-t_{3}
\end{array}\right)\left(\begin{array}{ccc}
1\\
 & -\frac{1}{2} & -\frac{\sqrt{3}}{2}\\
 & \frac{\sqrt{3}}{2} & -\frac{1}{2}
\end{array}\right)^{-j}\label{eq:6}
\end{equation}
where 
\begin{equation}
\mathbf{\delta}_{j}=C_{3}^{j}\frac{a}{\sqrt{3}}\left(\begin{array}{ccc}
0 & -1 & 0\end{array}\right)^{T},j=0,1,2\label{eq:7}
\end{equation}
and the sublattice vectors 
\[
\mathbf{\gamma}_{B1}=\frac{a}{\sqrt{3}}\left(\begin{array}{ccc}
\frac{\sqrt{3}}{2} & -\frac{1}{2} & 0\end{array}\right)^{T}
\]
\[
\mathbf{\gamma}_{B2}=\frac{a}{\sqrt{3}}\left(\begin{array}{ccc}
\frac{\sqrt{3}}{2} & \frac{1}{2} & 0\end{array}\right)^{T}
\]
\begin{equation}
\mathbf{\gamma}_{Mg}=c\left(\begin{array}{ccc}
0 & 0 & \frac{1}{2}\end{array}\right)^{T}\label{eq:8}
\end{equation}
where $a=3.086$ $\textrm{Å}$ is the lattice constant in the x-y
plane, and $c=3.524$ $\textrm{Å}$ is the lattice constant along the
$z-$direction. Here, the zeroes in the matrices are defined implicitly.

The second part of the Hamiltonian is given by 
\begin{equation}
H^{B,p_{z}}=\sum_{k}^{1BZ}c_{\mathbf{k},B,p_{z}}^{\dagger}h_{p_{z}}(\mathbf{k})c_{\mathbf{k},B,p_{z}}\label{eq:9}
\end{equation}
where $c_{\mathbf{k},B,p_{z}}^{\dagger}=\left(\begin{array}{cc}
c_{\mathbf{k},\gamma_{B1},p_{z}}^{\dagger} & c_{\mathbf{k},\gamma_{B2},p_{z}}^{\dagger}\end{array}\right)$, with
\begin{equation}
h_{p_{z}}(\mathbf{k})=\left(\begin{array}{cc}
h_{B1-B1,p_{z}}(\mathbf{k}) & h_{B1-B2,p_{z}}(\mathbf{k})\\
h_{B2-B1,p_{z}}(\mathbf{k}) & h_{B2-B2.p_{z}}(\mathbf{k})
\end{array}\right)\label{eq:10}
\end{equation}
and 
\begin{equation}
h_{B1-B1,p_{z}}(\mathbf{k})=E_{p_{z}}+t_{p_{z},z}2\cos(\mathbf{k}.\mathbf{a_{3}})\label{eq:11}
\end{equation}
\begin{equation}
h_{B2-B2,p_{z}}(\mathbf{k})=h_{B1-B1,p_{z}}(\mathbf{-k})\label{eq:12}
\end{equation}
\begin{equation}
h_{B1-B2,p_{z}}(\mathbf{k})=h_{B2-B1,p_{z}}^{\dagger}(\mathbf{k})=\sum_{j=0,1,2}\exp(-i\mathbf{\delta_{j}}.\mathbf{k})t_{p_{z}}\label{eq:13}
\end{equation}

The parameter values considered for the tight-binding Hamiltonian are 
\begin{align}
E_{B,s,0} & =-1.68,E_{B,p_{x},p_{y},0}=3.68,E_{p_{z}}=0,t_{1}=-3.5,t_{2}=1.022,t_{3}=-2.66,t_{4}=3.8,t_{p_{z}}=-1.7\nonumber \\
 & t_{B,s,z}=-0.085,t_{B,p_{x}p_{y},z}=-0.089,t_{p_{z},z}=1\label{eq:15}
\end{align}
in units of eV. The resulting bandstructure agrees well with \textit{ab initio}
calculations close to the Fermi energy. 

\subsection{Interaction effects}

The interacting part of the Hamiltonian is given by 
\begin{equation}
\frac{1}{2}\sum_{q}V(q)\rho(q)\rho(-q)\label{eq:16}
\end{equation}
with the Coulomb interaction
\begin{equation}
V(q)=\frac{e^2}{\epsilon_{0}\epsilon_{\infty}q^2}=\frac{U}{q^{2}}\label{eq:17}
\end{equation}
where momentum $q$ is measured in $\textrm{Å}^{-1}$ and $U$ is treated as a tunable
parameter, 
that we estimate by comparing the dynamical charge susceptibility
in the RPA, computed for a given interaction strength, directly with the energy-loss function determined using the GPAW open-source python package \cite{mortensen_gpaw_2024}, which gives a plasmon peak close to 3 eV. For the purpose of our calculations, we use $U=2$ $\rm{eV}\rm{\AA}$. We find that the low-energy spectrum is unaffected by the choice of the parameter $U$.

The charge density 
\begin{equation}
\rho(q)=\sum_{\mathbf{k},s}c_{\mathbf{k},s}^{\dagger}c_{\mathbf{k}+\mathbf{q},s}\label{eq:18}
\end{equation}
where $s=\{B_{i},s_{j}\}$, with $B_{i}=\{B_{1}, B_{2}\}$ referring to the B atom
and $s_{j}=\{s, p_{x}, p_{y}\}$ to the relevant orbital. 

Diagonalizing the noninteracting part of the Hamiltonian gives 
\begin{equation}
H=\sum_{k,a}^{1BZ} c_{k,a}^{\dagger}\epsilon_a(k)c_{k,a}\label{eq:19}
\end{equation}
where
\begin{equation}
c_{k,a}=\sum_{s}U_{sa}^{*}(k)c_{k,s}\label{eq:20}
\end{equation}
and $a$ refers to the band indices and $s$ to the orbital indices. There are eight bands, which
can be classified into six $\sigma-$bands originating from the
boron $s,p_{x},p_{y}$ orbitals, which disperse primarily in the x-y
plane and two $\pi-$bands originating from the boron $p_{z}$
orbitals, dispersing primarily in the z-direction. 

In the band basis, the charge density can be decomposed as
\begin{equation}
\rho(q)=\sum_{sab,\mathbf{k}}U_{sa}^{*}(\mathbf{k})U_{sb}(\mathbf{k}+\mathbf{q})c_{a}^{\dagger}(\mathbf{k})c_{b}(\mathbf{k}+\mathbf{q})=\sum_{ab}\rho_{ab}(\mathbf{q})\label{eq:21}
\end{equation}
where 
\begin{equation}
\rho_{ab}(\mathbf{q})=\sum_{s,\mathbf{k}}U_{sa}^{*}(\mathbf{k})U_{sb}(\mathbf{k}+\mathbf{q})c_{a}^{\dagger}(\mathbf{k})c_{b}(\mathbf{k}+\mathbf{q})\label{eq:22}
\end{equation}

\subsection{Charge susceptibility}

Here, we summarize our calculations for the charge susceptibility,
which describes the response of the total charge density to a potential
coupling to the total charge density.

The non-interacting charge susceptibility is given by 
\begin{equation}
\chi^{0}(q,\omega)=\frac{2}{N}\sum_{rsab,k}U_{ra}(k+q)U_{sa}^{*}(k+q)U_{sb}(k)U_{rb}^{*}(k)\left(\frac{f(\varepsilon_{b}(k))-f(\varepsilon_{a}(k+q))}{\omega+\varepsilon_{b}(k)-\varepsilon_{a}(k+q)+i\delta}\right)\label{eq:23}
\end{equation}
with $\delta\rightarrow0^{+}$, where $r,s$ refers to the orbital
indices and $a,b$ to the band indices, $N$ is the number of $k-$points
summed over and $f(\varepsilon)$ is the Fermi-Dirac function for energy $\varepsilon$. For
the purpose of our calculations, we use a N$\times$N$\times$N
grid of $k-$points, which are uniformly distributed over the first
Brillouin zone, where N=512. The temperature is set to zero and a small Lorentzian broadening
of $\delta=0.01$ $\rm{eV}$ is applied. 
The total non-interacting susceptibility can be written as
\begin{equation}
\chi^{0}(q,\omega)=2\sum_{ij}\chi_{ij}^{0}(q,\omega)=\sum_{abcd}\delta_{ad}
\delta_{bc}\chi_{ab,cd}^{0}\label{eq:25b}
\end{equation}
where $i,j=\{s, p_{x},p_{y},p_{z}\}$ refers to the B orbital indices and
$a,b,c,d=\{\sigma,\pi\}$ refer to the band indices. 
%The full charge susceptibility within the RPA is given by 
%\begin{equation}
%\chi^{RPA}(q,\omega)=\frac{\chi^{0}(q,\omega)}{1-V(q)\chi^{0}(q,\omega)}\label{eq:24}
%\end{equation}

We can also define a susceptibility matrix where each element describes
the response of a component of the charge density (in band basis)
to a potential coupling to any other component, as 
\begin{equation}
\chi_{ab,cd}(q,i\omega_{n})=-\frac{1}{N}\int_{0}^{\beta}d\tau\exp(i\omega_{n}\tau)\left(\langle\rho_{ab}(q,\tau)\rho_{cd}(-q)\rangle-\langle\rho_{ab}(q,\tau)\rangle\langle\rho_{cd}(-q)\rangle\right)\label{eq:25}
\end{equation}
The susceptibility is obtained analytically continuing $i\omega_{n}\rightarrow\omega+i\delta,\delta\rightarrow0^{+}$. 

The noninteracting susceptibility decomposed into band components
is 
\begin{equation}
\chi_{ab,cd}^{0}(q,\omega)=\delta_{ad}\delta_{bc}\frac{2}{N}\sum_{rs,k}U_{ra}(k+q)U_{sa}^{*}(k+q)U_{sb}(k)U_{rb}^{*}(k)\left(\frac{f(\varepsilon_{b}(k))-f(\varepsilon_{a}(k+q))}{\omega+\varepsilon_{b}(k)-\varepsilon_{a}(k+q)+i\delta}\right)\label{eq:26}
\end{equation}
where the bands are decoupled in the non-interacting system. 

The interaction term may be rewritten as 
\begin{equation}
\sum_{abcd}V_{ab,cd}(q)\rho_{ab}(q)\rho_{cd}(-q)\label{eq:27}
\end{equation}
where $V_{ab,cd}(q)=V(q)$ for all $a,b,c$ and $d$. 

For our analysis, we
take interaction effects into account through the random phase approximation (RPA). 

Within the RPA, the susceptibility takes a matrix form given by 
\begin{equation}
\mathbf{\chi}^{RPA}(q,\omega)=\mathbf{\chi^{0}}(q,\omega)[\mathbf{I}-\mathbf{V}(q)\mathbf{\chi^{0}}(q,\omega)]^{-1}\label{eq:28}
\end{equation}
where $\mathbf{I}$ is the identity matrix and the multiplication
and inversion are matrix operations here. 

In addition to interaction effects, we also take into account the experimental geometry and photon polarization.
For $\pi-$polarization
of the incoming photons, and for the convention used in the experiment,
incident angle $\theta_{in}=0^{\circ}$ corresponds to normal incidence,
with polarization vector parallel to the surface of the sample, and
incident angle $\theta_{in}=90^{\circ}$ corresponds to grazing incidence,
with polarization vector along the $c-$axis. Therefore, we focus
on the component $\chi^{RPA}_{p_{z},p_{z}}$ for grazing incidence, and
the component $\chi^{RPA}_{p_{x},p_{x}}$ or $\chi^{RPA}_{p_{y},p_{y}}$(depending
on the direction of the in-plane component of the momentum, $q_{\Vert}$)
for normal incidence, using appropriate sinusoidal weight factors.
In practice, we compare the RIXS spectrum directly with the weighted combination of different orbital components of the RPA susceptibility, which we refer to as $\chi$ below
\begin{equation}
\chi_{\Gamma M}=\chi_{p_{y},p_{y}}^{RPA}\cos^{4}\theta_{in}+\chi_{p_{z},p_{z}}^{RPA}\sin^{4}\theta_{in}
+\left(\chi_{p_{y},p_{z}}^{RPA}+\chi_{p_{z},p_{y}}^{RPA}\right)\cos^{2}\theta_{in}\sin^{2}\theta_{in}\label{eq:31}
\end{equation}
for in-plane momentum along the $\Gamma M$ direction (which can
be chosen to be along the cartesian $y-$axis), and 
\begin{equation}
\chi_{\Gamma K}=\chi_{p_{x},p_{x}}^{RPA}\cos^{4}\theta_{in}+\chi_{p_{z},p_{z}}^{RPA}\sin^{4}\theta_{in}
+\left(\chi_{p_{x},p_{z}}^{RPA}+\chi_{p_{z},p_{x}}^{RPA}\right)\cos^{2}\theta_{in}\sin^{2}\theta_{in}\label{eq:32}
\end{equation}
for in-plane momentum along the $\Gamma K$ direction (which can be
chosen to be along the cartesian $x-$axis). 

\subsection{Orbital selectivity toy model}
\label{orbitalselect}

To understand why RIXS can be sensitive to intra-band particle-hole excitations that are difficult to detect by other techniques, let us analyze the consequences of the interactions under the RPA. The charge susceptibility under the RPA is given by
\begin{equation}
    \chi(q,\omega) = \frac{\chi_0(q,\omega)}{1 - V(q) \chi_0(q,\omega)}.
\end{equation}
Here, $V(q) \sim\frac{1}{q^2}$ is the Coulomb interaction and $\chi_0(q,\omega)$ is the non-interacting susceptibility, which consists purely of intra-band and inter-band particle-hole excitations.

The cross-section of spectroscopic probes is related to the imaginary part of the susceptibility
\begin{equation}
    \Im\chi = \frac{\Im \chi_0}{(1 - V \Re \chi_0)^2 + (V \Im \chi_0)^2} \label{eqimrpa}
\end{equation}
From \eqref{eqimrpa}, we see that $\Im\chi$ is nonzero only if $\Im\chi_0$ is nonzero or if the denominator vanishes. The latter situation gives rise to collective excitations, namely plasmons. The former indicates that the elementary particle-hole excitations that constitute the non-interacting response are still present in the interacting response function. 

Considering the limit of small $q$ and large $V$, \eqref{eqimrpa} simplifies to
\begin{equation}
    \Im\chi \approx \frac{\Im \chi_0}{V^2 \abs{\chi_0}^2} \sim \frac{q^4}{\abs{\chi_0}^2} \Im\chi_0. \label{eqimrpa_approx}
\end{equation}
We therefore see that the elementary excitations present in $\chi_0$ are suppressed by a factor of $q^4$ in the charge susceptibility $\chi$. This is the chief reason why elementary particle-hole excitations are not detected by techniques such as EELS, which probes the total charge response. RIXS, on the other hand, has a more complicated cross-section with orbital sensitivity that is strongly dependent on the scattering geometry. As an illustration, suppose the total charge density can be decomposed into that of two orbitals $a$ and $b$. The charge susceptibility can be decomposed into a $2 \times 2$ matrix corresponding to the response of the components to fields that couple to each of them. Assuming a diagonal non-interacting susceptibility, for simplicity, we have 
\begin{align}
    \boldsymbol\chi^0 &= \begin{pmatrix} \chi_a^0 & 0 \\ 0 & \chi_b^0 \end{pmatrix} \\
   % \boldsymbol\chi &= \boldsymbol\chi^0 \qty[ \mathbf{I} - \begin{pmatrix} V & V \\ V & V \end{pmatrix} \boldsymbol\chi^0]^{-1}
\end{align}
Under the RPA, from Eq. \eqref{eq:28}, the diagonal components of the interacting susceptibility $\boldsymbol\chi$ have an imaginary part no longer suppressed by powers of $q$. For instance,
\begin{equation}
\Im \chi_{aa} = \frac{\abs{\chi_b^0}^2 \Im \chi_a^0}{\abs{\chi_a^0+\chi_b^0}^2} + \frac{\abs{\chi_a^0}^2 \Im \chi_b^0}{\abs{\chi_a^0+\chi_b^0}^2} + \mathcal{O}\qty(\frac{1}{V}) \label{eq:imchiaa}
\end{equation}
Thus, any degree of orbital selectivity significantly elevates the intensity of the elementary particle-hole excitations in the response function.

As a demonstration of this effect, we consider a non-interacting Fermi gas with two identical bands such that $\chi_a^0 = \chi_b^0$. The total charge susceptibility calculated by the RPA (\figref{fig:toyrpa}{b}) shows almost all the spectral weight is in the plasmon, and the Lindhard continuum has minimal spectral weight except where plasmon intersects the continuum and becomes Landau damped. By contrast, the partial charge susceptibility (\figref{fig:toyrpa}{c}) displays both the plasmon and the particle-hole continuum with significant spectral weight. This confirms the contrasting behaviors of \eqref{eqimrpa_approx} and \eqref{eq:imchiaa}.

\newpage

\begin{figure*}
    \centering
    \includegraphics[width=120mm]{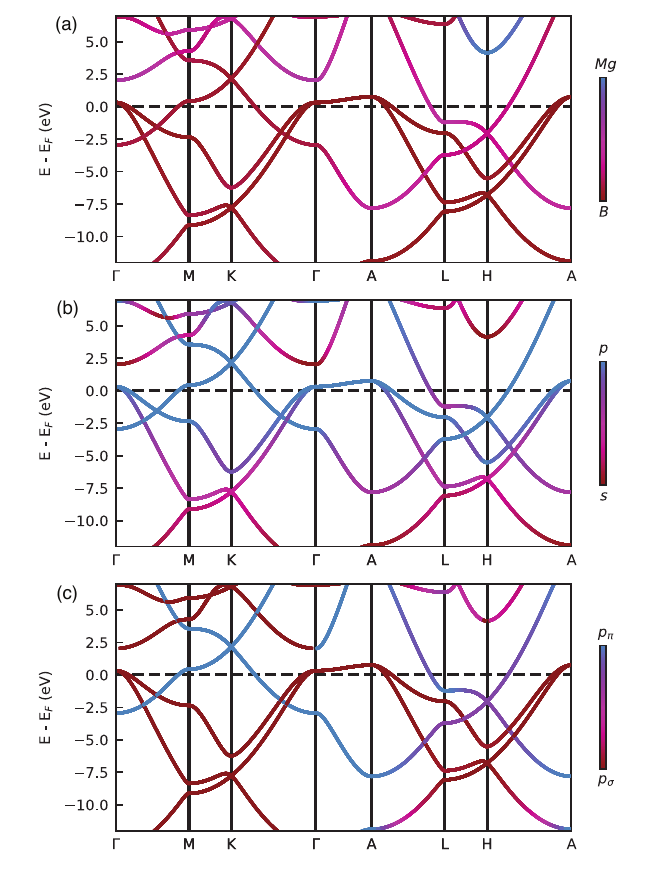}
    \caption{Atomic (a), orbital (b), and $\sigma$ and $\pi$ $p$-orbital character (c) projected band structure of MgB$_2$.}
    \label{fig:all_bs_suppl}
\end{figure*}

\begin{figure*}
    \centering
    \includegraphics[width=70mm]{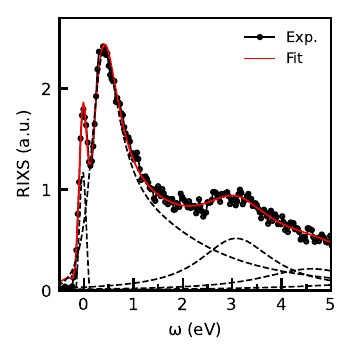}
    \caption{Example of a voigt fit of our RIXS data, showing an elastic, lower energy, and higher energy primary components.}
    \label{fig:voigt_fit}
\end{figure*}

\begin{figure*}
    \centering
    \includegraphics[width=70mm]{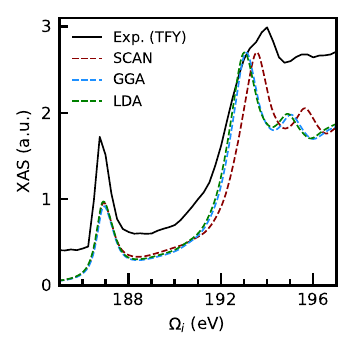}
    \label{fig:XAS_xc_comp}
\end{figure*}

\begin{figure*}
    \centering
    \includegraphics[width=120mm]{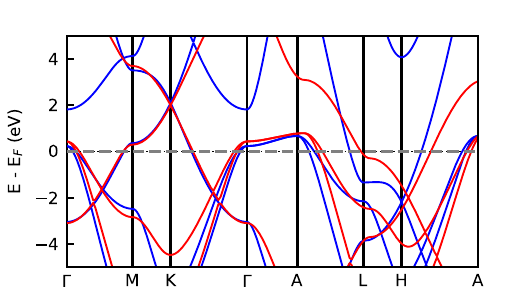}
    \caption{A comparison of the bandstructure obtained from the simplified tight-binding model (solid red lines) with that obtained from the Quantum ESPRESSO DFT calculation (solid blue lines).}
    \label{fig:gpaw_bs}
\end{figure*}

\begin{figure*}
    \centering
    \includegraphics{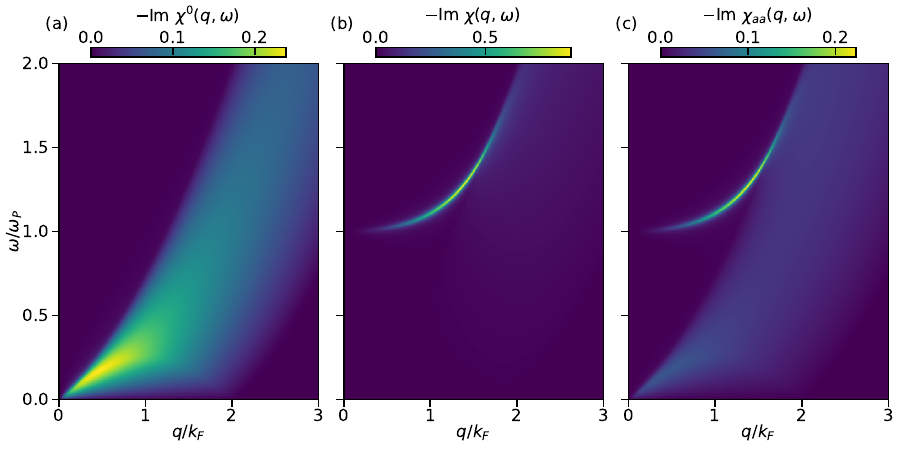}
    \caption{(a) Non-interacting Lindhard function, (b) total susceptibility, (c) partial susceptibility of a Fermi gas with two identical bands.}
    \label{fig:toyrpa}    
\end{figure*}

\end{document}